\documentclass[aps,pra,preprint,amsmath,showpacs,groupedaddress,endfloats*]{revtex4-1}
\usepackage{graphicx}
\usepackage[breaklinks=true]{hyperref}

\begin{document}

\title{Surface-assisted ultralocalization in nondiffracting beams}

\author{Juan J. Miret}
\affiliation{Departamento de \'Optica, Universidad de Alicante, P.O. Box 99, Alicante, Spain.}

\author{Carlos J. Zapata-Rodr\'{\i}guez}
\email{carlos.zapata@uv.es}
\affiliation{Departamento de \'Optica, Universidad de Valencia, Dr. Moliner 50, 46100 Burjassot, Spain.}

\date{\today}

\begin{abstract}
We present a family of localized radiation modes in multilayered periodic media,
where in-phase superposition of p-polarized waves leads to radiative confinement around the beam axis.
Excitation of surface plasmon polaritons yields an enhanced localization normally to the interfaces. 
On the other hand, the spectral filtering induced by the presence of bandgaps allows to achieve transverse superresolution on the interfaces.
Subwavelength beamwidths along an infinitely long distance might potentially be obtained. 
\end{abstract}


\maketitle

\section{\label{sec01} Introduction}
 
Bessel beams \cite{Durnin87,Durnin87b,Recami08} are a family of nondiffracting waves propagating in dielectric media along for instance the $z$-axis with a homogeneous spatial variation $\exp \left( i \beta z \right)$ but transversally localized around its \emph{focus}, say $r = 0$.
For a zero-order Bessel beam, the transverse pattern is written in terms of the radially-symmetric function $J_0 \left( k_\perp r \right)$.
They are interpreted as a suitable superposition of plane waves all having a wavevector which projected onto the propagation axis yields the characteristic propagation constant $\beta > 0$ of the Bessel beam \cite{Indebetouw89,Fagerholm96}.
A phase matching condition at focus is additionally satisfied in order to provide the best spatial localization of the wave in the vicinities of such an axis.
The FWHM of the Bessel beam is $3.04\ k_\perp^{-1}$, which increases along with the propagation constant in accordance with the dispersion equation $k_\perp^2 + \beta^2 = n^2 k_0^2$, being $n$ is the index of refraction of the medium and $k_0 = 2 \pi / \lambda_0$ the wavenumber in vacuum.
Therefore the minimum beamwidth, roughly given by $\lambda_0 / 2 n$, is reached in the stationary-wave limit $\beta = 0$.

This sort of solutions of the Helmholtz equation may be found also in stratified media if the unit vector ($\mathbf{n}$) normal to the interfaces lies along the $z$-axis leading to normal incidence of the wavefield \cite{Mugnai00,Longhi04,Novitsky06,Zapata08b}.
However, out-of-plane excitation cannot support the invariant propagation of Bessel beams \cite{Zapata10a}.
For example, assuming the medium is periodic, the elements of any wave superposition are necessarily Bloch modes leading to a significant asymmetry in the resultant pattern.
Provided the projection $\beta$ of the pseudomoment along a direction perpendicular to the periodicity of the medium coincides for all Bloch components we may have a nondiffracting localized beam if, additionally, the phase matching condition is satisfied \cite{Miret08}.
With dielectric slabs, we encounter a higher localization of the field for decreasing values of the propagation constant $\beta$. 
This is also reported in two-dimensional photonic crystals \cite{Christodoulides03,Manela05}, being observed not only under subdiffractive light propagation \cite{Staliunas06} but much more generally.
 
In this paper we study the formation of localized waves which propagation is resonantly sustained on a given interface from a 1D periodic metallo-dielectric medium.
The surface resonance arises in transverse-magnetic waves ($\mathbf{H} \perp \mathbf{n}$) provided the sign of the dielectric constant changes abruptly at both sides of the interface.
The excitation of such surface plasmons polaritons (SPPs) are attained at comparatively high values of $\beta$, however leading to a subwavelength beam size.
In spite of such an extreme wave localization the filamentation of the excited field in the linear medium is ideally maintained for infinitely-long distances. 

\section{\label{sec02} In-plane propagation of diffraction-free beams in metallo-dielectric layered media}

\begin{figure}
\centering
\includegraphics[width=6.5cm]{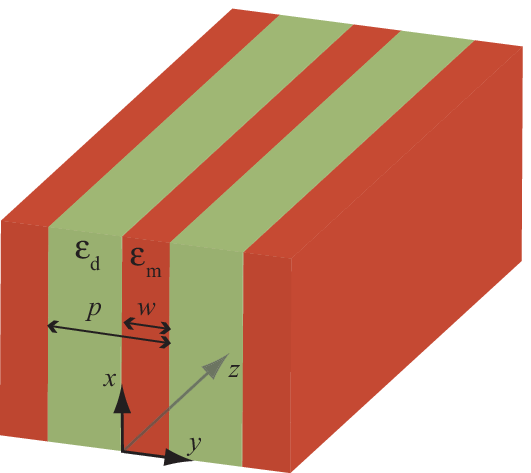}
\caption{Schematic geometry of the planar-nanolayer-based medium.} 
\label{fig01}
\end{figure}

Let us consider a monochromatic nondiffracting beam propagating in a multilayered medium.
The $y$-axis is set such that it is perpendicular to the surfaces separating the metallic media and the adjacent dielectric media, so that $\mathbf{n} = \hat{\mathbf{y}}$.
In Fig.~\ref{fig01} we show schematically the multilayered system.
The width of a metallic slab is $w$, an element periodically replicated at a distance $p$ along the $y$-axis.
We also assume that beam propagation is directed along the $z$-axis, so that we may cast the electromagnetic fields as
\begin{subequations}
\begin{eqnarray}
 \mathbf{E} (x,y,z,t) = \mathbf{e} (x,y) \exp \left( i \beta z - i \omega t \right) , \\
 \mathbf{H} (x,y,z,t) = \mathbf{h} (x,y) \exp \left( i \beta z - i \omega t \right) ,
\end{eqnarray}
\label{eq01}
\end{subequations}
being $\omega$ the frequency of the monochromatic radiation.
The homogeneity of the wave field in the coordinate $z$ is explicitly parametrized in terms of the propagation constant $\beta$.
More specifically, we study TM waves where $\mathbf{H}$ exists only onto planes parallel to the metal-dielectric interfaces, and therefore which component $h_y$ vanishes.
Later on we additionally consider wave confinement around the origin $(x,y) = (0,0)$ in any given transverse plane.

The Maxwell's equations provide some relations between the transverse fields $\mathbf{e}$ and $\mathbf{h}$.
The electric field $\mathbf{e}$ may be derived from $\mathbf{h}$ by means of the equation
$\nabla \times \mathbf{H} = -i \omega \epsilon_0 \epsilon \mathbf{E}$,
where $\epsilon(y)$ is the relative dielectric constant of the foliar structure.
Thus we have
\begin{subequations}
\begin{eqnarray}
 e_x &=& i \left( \omega \epsilon_0 \epsilon \right)^{-1} \partial_y h_z , \label{eq02a} \\
 e_y &=& i \left( \omega \epsilon_0 \epsilon \right)^{-1} \left( i \beta h_x - \partial_x h_z \right) , \\
 e_z &=& - i \left( \omega \epsilon_0 \epsilon \right)^{-1} \partial_y h_x . \label{eq02c}
\end{eqnarray}
\label{eq02}
\end{subequations}
From the equation $\nabla \left( \mu_0 \mathbf{H} \right) = 0$ we also find
\begin{equation}
 h_z = i \beta^{-1} \partial_x h_x ,
\label{eq03}
\end{equation}
so that $h_x$ is the scalar wavefield from which we may describe the nondiffracting beam unambiguously.
Let us point out that alternate routes for the description of electromagnetic diffraction-free beams may be found elsewhere \cite{Bouchal95b,Bouchal98,Paakkonen02}.

The two-dimensional Helmholtz equation
\begin{equation}
 \left( \partial_x^2 + \partial_y^2 + \omega^2 \mu_0 \epsilon_0 \epsilon - \beta^2 \right) h_x = 0 .
\label{eq04}
\end{equation}
may be of help in order to find any localized solution of the field $h_x$.
However it is preferable an alternative approach.
Based on the Bloch theorem \cite{Joannopoulos08}, we may describe the propagating wavefield $h_x$ as a superposition of Bloch modes having the form
\begin{equation}
 h_x (x,y) = \sum_K \int_{-\infty}^{\infty} a_K h_K (y) \exp \left( i k_x x + i K y \right) d k_x ,
\label{eq05}
\end{equation}
where $k_x$ is the (real) component of a wave vector along the $x$-axis, and $K$ is the so-called Block wave number, which generally is multivalued for a given $k_x$.
Moreover, $h_K (y + p) = h_K (y)$ is a periodic function, which is normalized as $h_K (0) = 1$ for convenience, and $a_K$ is simply a field amplitude.

Assuming that sources are sufficiently far from the focal region of the localized beam, the role of evanescent waves \cite{Ramakrishna04} is negligible in our analysis and, therefore, they are disregarded setting $a_K = 0$ if $\mathrm{Im}(K) \neq 0$.
In this sense we also neglect material losses imposing that the relative dielectric constant of the medium is a real parameter of the problem [$\mathrm{Im}(\epsilon) = 0$].
We point out that this is not strictly true in a realistic problem; plasma-like media with $\epsilon < 0$ are necessarily dispersive since energy density considerations lead to the condition $d (\omega \epsilon) / d \omega > 0$, and the presence of dispersion in general signifies dissipation of energy \cite{Landau84}.
However, it is possible to neglect the absorption in a transparency window where the imaginary part of $\epsilon$ is very small in comparison with its real part.

In the case of the stratified medium shown in Fig.~\ref{fig01}, the periodic function is conveniently written as the addition of two independent terms, $h_K = h_K^+ + h_K^-$.
Using the column vector
\begin{equation}
 \mathbf{h}_K
 =
 \left[
  \begin{array}{c}
   h_K^+ \\
   h_K^-
  \end{array}
 \right] ,
\end{equation}
we may write \cite{Yeh88}
\begin{equation}
 \mathbf{h}_K (y) = \exp \left( - i K y \right) \mathbf{T}_\alpha (y - y_\alpha) \mathbf{v}_\alpha \mbox{, for } y \in R_\alpha ,
\label{eq06}
\end{equation}
where $\alpha$ is an integer that refers to a unique slab.
The domains $R_\alpha \equiv (y_{\alpha - 1},y_\alpha]$ where $y_0 = 0$, $y_1 = w$ and $y_{\alpha + 2} = y_\alpha + p$; for instance, $R_\alpha$ is associated with a region where the medium is metallic when $\alpha$ is an odd number.
In Eq.~(\ref{eq06}), the translation matrix
\begin{equation}
 \mathbf{T}_\alpha(y) = \left[
             \begin{array}{cc}
              \exp \left( i k_{y \alpha} y \right) & 0 \\
              0                                    & \exp \left( - i k_{y \alpha} y \right)
             \end{array}
                        \right] ,
\label{eq07}
\end{equation}
and the column vector $\mathbf{v}_\alpha = [a_\alpha, b_\alpha]$ 
giving some constant amplitudes. 
Note that $a_\alpha$ is independent of $a_K$ given in Eq.~(\ref{eq05}) and they should not be confused.
Also, 
\begin{equation}
 k_{y \alpha} = \left\{
 \begin{array}{cc}
  \sqrt{\left( \omega / c \right)^2 \epsilon_\alpha - k_\parallel^2}   ,\  & \left(\omega / c \right)^2 \epsilon_\alpha \ge k_\parallel^2 \\
  i \sqrt{k_\parallel^2 - \left( \omega / c \right)^2 \epsilon_\alpha} ,\  & \left(\omega / c \right)^2 \epsilon_\alpha <   k_\parallel^2
 \end{array}
 \right.
\label{eq08}
\end{equation}
where $\epsilon_1 = \epsilon_m$ and $\epsilon_2 = \epsilon_d$ are the relative dielectric constants of the metallic medium and the dielectric material, respectively.
For other slabs we use the recurrence relation $\epsilon_{\alpha + 2} = \epsilon_\alpha$; here $\epsilon_\alpha$ at $\alpha = 0$ does not refers to its value in vacuum but $\epsilon_d$ and it is not necessarily the unity.
In Eq.~(\ref{eq08}) the real-valued wave vector onto a plane parallel to the metallo-dielectric interface has a squared modulus $k_\parallel^2 = k_x^2 + \beta^2$.
Finally since material losses are neglected, the dielectric constant of the plasma-like material $\epsilon_m < 0$ and that of the insulator $\epsilon_d >0$ are also real-valued parameters.

We also take advantage of spatial symmetries from the layered system in Fig.~\ref{fig01}.
In particular, we use that the relative dielectric constant is a real and even function $\epsilon(-y') = \epsilon(y')$, being $y' = y - y_{\alpha - 1/2}$ a spatial coordinate centered in the middle point of the domain $R_\alpha$, $y_{\alpha - 1/2} = \left( y_\alpha + y_{\alpha - 1} \right) / 2$.
Given a wavefield $h_x (x,y')$ that satisfies Eq.~(\ref{eq04}), the fields $h_x^* (x,y')$ and $h_x (x,-y')$ are also solutions of the wave equation.
From the series expansion of Eq.~(\ref{eq05}), therefore, we infer that $h_K(-y')$ and $h_K^*(y')$ are periodic functions both associated with the same Bloch (real) number $-K$, so that there should be a constant of proportionality between them.
Provided $\varphi$ is the argument of $h_K(y'=0)$ we find that
\begin{equation}
 h_K(-y') \exp (-i \varphi) = \left[ h_K(y') \exp (-i \varphi) \right]^* .
\label{eq22}
\end{equation}
As a consequence $h_K$ is hermitian under the normalization $h_K (y' = 0) = 1$.
However this is not necessarily true when $h_K (y = 0) = 1$.
In the later case it yields for instance that the amplitude of the periodic function $|h_K|$ evaluated on every surface $y = y_\alpha$ reaches the unit.

The continuity at the interfaces of the wavefields $h_x$ given in Eq.~(\ref{eq05}), $h_z$, $e_x$ and $e_z$ [that is continuity of $\epsilon^{-1} \partial_y h_x$ as inferred from Eqs.~(\ref{eq02a}) and (\ref{eq02c})], together with the periodicity of $h_K$ determine the waveform of each Bloch mode and its dispersion behaviour.
Using the transfer matrix elements for layered media \cite{Yeh88} we may obtain in a rather simple way the dispersion equation.
Let us give the corresponding analysis briefly.
Experienced readers may ignore the rest of the present section thus going to Sec.~\ref{sec03} directly.

\subsection{Dispersion equation and isofrequency curves}

Eq.~(\ref{eq06}) provides the complex waveform of $\mathbf{h}_K$ appropriately.
However, in order to simplify the application of continuity conditions of the fields, we also establish an equivalent definition,
\begin{equation}
 \mathbf{h}_K (y) = \exp \left( - i K y \right) \mathbf{T}_\alpha (y - y_{\alpha - 1}) \mathbf{v}'_\alpha
\end{equation}
at $y \in R_\alpha$, where the new column vector $\mathbf{v}'_\alpha = [a'_\alpha, b'_\alpha]$.
Note that 
\begin{equation}
 \mathbf{v}'_\alpha = \mathbf{P}_\alpha \mathbf{v}_\alpha ,
\label{eq12}
\end{equation}
the matrix $\mathbf{P}_\alpha = \mathbf{T}_\alpha(y_{\alpha - 1} - y_\alpha)$ representing a translation between the boundaries of $R_\alpha$.
Thus the continuity of the p-polarized fields at $y = y_\alpha$ holds when
\begin{equation}
 \mathbf{D}_\alpha \mathbf{v}_\alpha = \mathbf{D}_{\alpha + 1} \mathbf{v}'_{\alpha + 1} ,
\label{eq13}
\end{equation}
where the transmission matrix
\begin{equation}
 \mathbf{D}_\alpha = \left[
             \begin{array}{cc}
              1                                    & 1 \\
              \frac{k_{y \alpha}}{\epsilon_\alpha} & -\frac{k_{y \alpha}}{\epsilon_\alpha}
             \end{array}
            \right] .
\end{equation}
Consequently, the field amplitudes corresponding to the slab labelled as $(\alpha + 1)$ may be determined by means of those amplitudes of the adjacent slab $\alpha$ as $\mathbf{v}_{\alpha + 1} = \mathbf{P}_{\alpha + 1}^{-1} \mathbf{D}_{\alpha + 1}^{-1} \mathbf{D}_\alpha \mathbf{v}_\alpha$.

On other hand, imposing the periodicity condition $h_K (y_{\alpha + 2}) = h_K (y_\alpha)$ we also find
\begin{equation}
 \mathbf{v}_{\alpha + 2} = \exp \left( i K p \right) \mathbf{v}_\alpha ,
\label{eq14}
\end{equation}
given as a direct application of the Bloch theorem.
As a consequence, considering the continuity of the fields at $y = y_\alpha$ given in (\ref{eq13}) and that at $y = y_{\alpha + 1}$, Eq.~(\ref{eq14}) leads to the matrix equation
\begin{equation}
 \mathbf{M}_\alpha \mathbf{v}_\alpha = \exp \left( i K p \right) \mathbf{v}_\alpha ,
\label{eq15}
\end{equation}
where Eq.~(\ref{eq12}) is also used at $(\alpha + 1)$ and $(\alpha + 2)$.
The computation of the matrix $\mathbf{M}_\alpha$ leads to
\begin{equation}
 \left[
  \begin{array}{cc}
   A_\alpha & B_\alpha \\
   C_\alpha & D_\alpha
  \end{array}
 \right] 
 = \mathbf{P}_{\alpha + 2}^{-1} \mathbf{D}_{\alpha + 2}^{-1} \mathbf{D}_{\alpha + 1} \mathbf{P}_{\alpha + 1}^{-1} \mathbf{D}_{\alpha + 1}^{-1} \mathbf{D}_\alpha .
\end{equation}
Periodicity in the medium renders $\mathbf{P}_{\alpha + 2} = \mathbf{P}_{\alpha}$ and $\mathbf{D}_{\alpha + 2} = \mathbf{D}_{\alpha}$ which yield further simplifications in the determination of $\mathbf{M}_\alpha$.
We leave the reader the long but straightforward evaluation of the elements of the matrix $\mathbf{M}_\alpha$.
In general $\mathbf{M}_{\alpha + 1} \neq \mathbf{M}_{\alpha}$; however $\mathbf{M}_{\alpha + 2} = \mathbf{M}_{\alpha}$ and, more importantly, the matrix $\mathbf{M}_{\alpha}$ is unimodular [$\mathrm{det}(\mathbf{M}_{\alpha}) = 1$] and its trace is independent of $\alpha$.

In agreement with Eq.~(\ref{eq15}), $\mathbf{v}_\alpha$ is an eigenvector of the matrix $\mathbf{M}_\alpha$ providing the eigenvalue $\exp \left( i K p \right)$.
Consequently the determinant of the matrix $\mathbf{M}_\alpha - \exp \left( i K p \right) \mathbf{I}$ vanishes, where $\mathbf{I}$ is the identity matrix.
The resultant equation may be cast simply as $\cos \left( K p \right) = \left( A_\alpha + D_\alpha \right) / 2$, where the unimodularity property of the matrix $\mathbf{M}_\alpha$, $A_\alpha D_\alpha - B_\alpha C_\alpha = 1$, should be employed.
Thus we have derived the dispersion equation, which explicitly reads 
\begin{eqnarray}
\label{eq09}
 \cos \left( K p \right) &=& \cos [k_{y d} (p-w)] \cos (k_{y m} w) \\
                         &-& \frac{\left( k_{y m}^2 \epsilon_d^2 + k_{y d}^2 \epsilon_m^2 \right)}{2 k_{y d} k_{y m} \epsilon_d \epsilon_m} \sin [k_{y d} (p-w)] \sin (k_{y m} w) \nonumber .
\end{eqnarray}
As expected, the dispersion equation (\ref{eq09}) is independent of the integer $\alpha$.
Our interest is focused in real values of $K$, for which the right-hand side of Eq.~(\ref{eq09}) should be comprised between $-1$ and $1$.

\begin{figure}
\centering
\includegraphics[width=7cm]{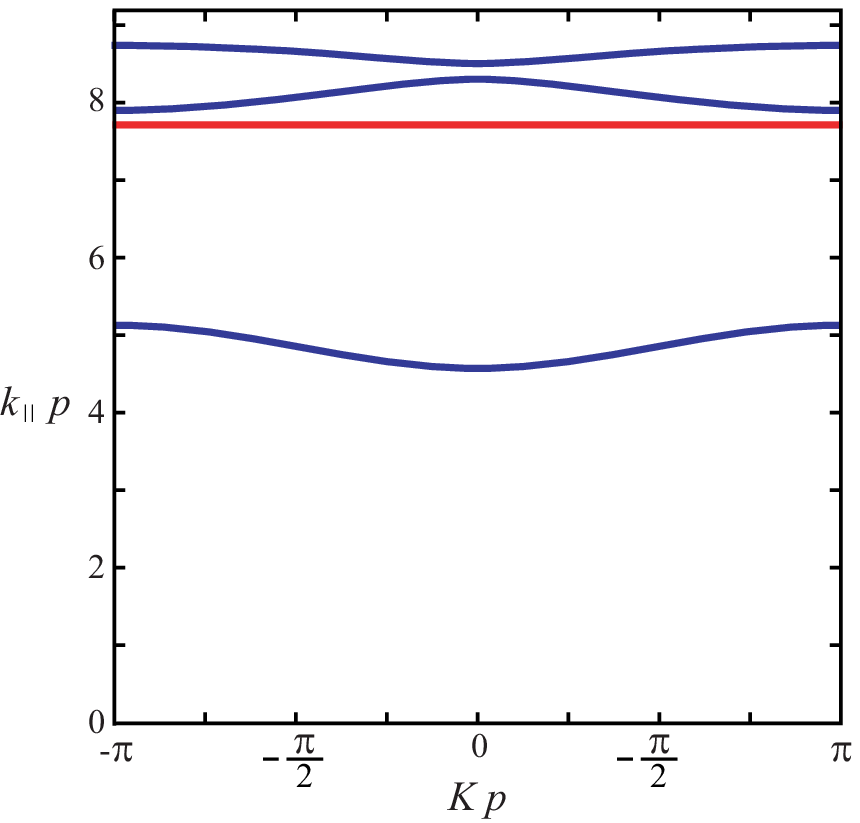}
\caption{Dispersion equation at $\omega = 3.4 \times 10^{15}\ \mathrm{rad/s}$ for a periodic media as represented in Fig.~\ref{fig01} with $w = 5 \times 10^{-8}\ \mathrm{m}$ and $p = 4.5 \times 10^{-7}\ \mathrm{m}$.
The red lines marks the boundary of homogeneous-wave and evanescent-wave regimes in the dielectric, $k_\parallel p = 7.711$.} 
\label{fig02}
\end{figure}

In Fig.~\ref{fig02} we plot the real-valued $K$ solutions of equation (\ref{eq09}) for a multilayer medium composed of thin metallic films of width $w = 50\ \mathrm{nm}$, separated a distance $p = 450\ \mathrm{nm}$, and embedded in a dielectric medium of $\epsilon_d = 2.25$.
At a frequency $\omega = 3.427\ \mathrm{fs}^{-1}$ (wavelength $\lambda_0 = 550\ \mathrm{nm}$ in the vacuum) our plasma-like medium has a relative dielectric constant $\epsilon_m = -15.0$.
A large bandgap in the interval $k_\parallel p \in (0,4.571)$ is followed by some two others in $(5.126,7.899)$ and $(8.301,8.503)$, together with the evanescent-wave regime at $k_\parallel p > 8.737$.
As a consequence, real values of $\beta$ are limited by $\beta_{max} = 8.737 / p$ in our example ($\beta_{max} = 19.42\ \mu\mathrm{m}^{-1}$).
This boundary value is reached at $K = \pm \pi / p$ rather than at a zero value \cite{Kuzmiak97}. 
Let us point out that, if the superposition shown in Eq.~(\ref{eq05}) is such that $k_\parallel$ is higher than the cut-off frequency 
\begin{equation}
 \beta_c = \sqrt{\epsilon_d} \omega / c ,
\label{eq18}
\end{equation}
occurring if $k_\parallel p > 7.711$, the wavefields are all of evanescent nature in the metal and in the dielectric; here $\beta_c = 17.14\ \mu\mathrm{m}^{-1}$.
The cut-off frequency is plotted with a horizontal red line in Fig.~\ref{fig02}.
Interestingly, these evanescent waves may arise at values of $k_\parallel p$ such that Eq.~(\ref{eq09}) gives a Bloch wave number with $\mathrm{Im}(K) = 0$.
Thus, the evanescent waves are resonantly coupled leading to propagating Bloch-type constituents of the diffraction-free wavefield $h_x (x,y)$.
Nondiffracting beams with a propagation constant $\beta \in (\beta_c, \beta_{max})$ are wave fields of this kind.

\begin{figure}
\centering
\includegraphics[width=8cm]{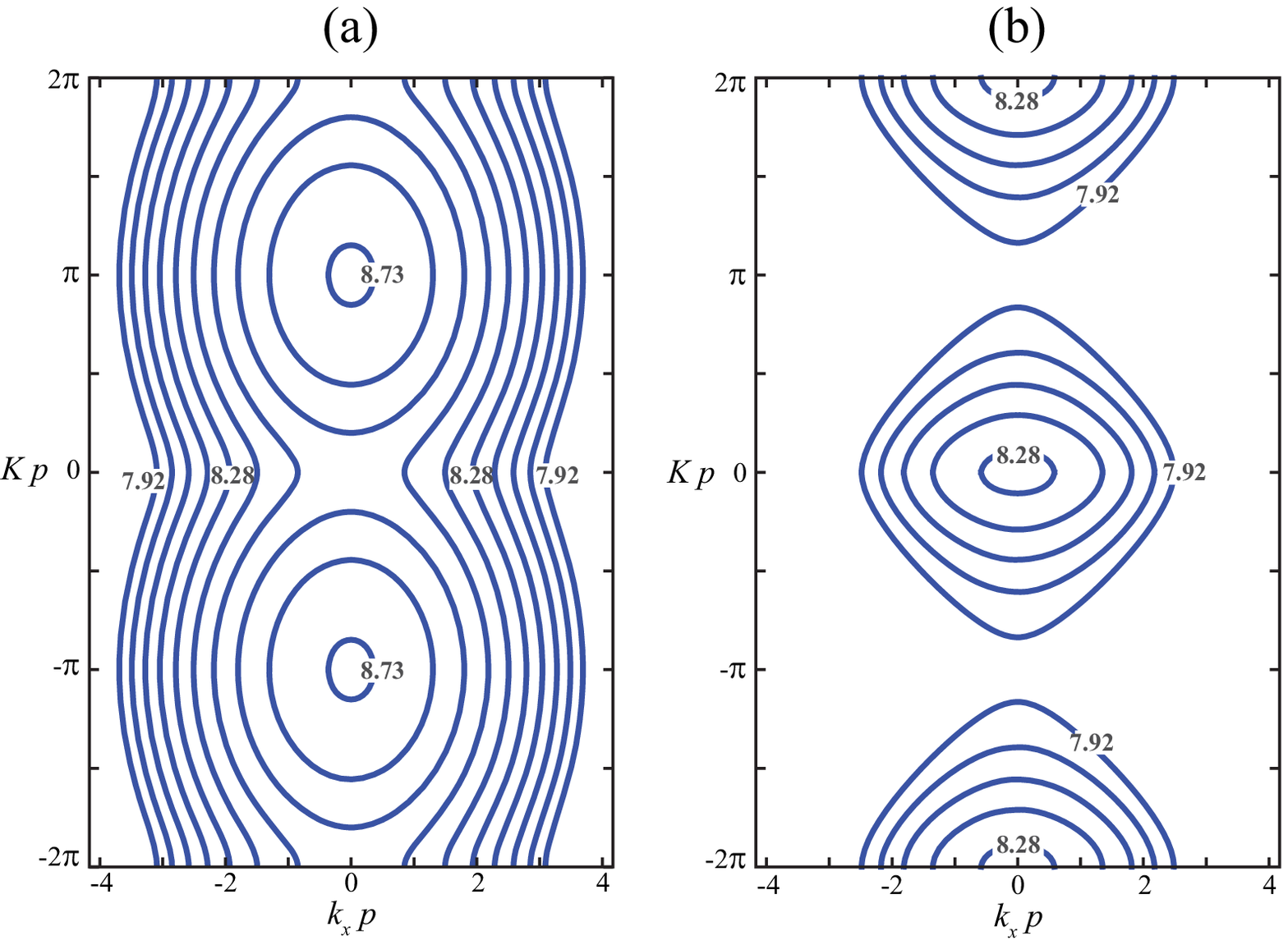}
\caption{Isofrequency curves at different propagation constants $\beta_c < \beta < \beta_{max}$ are shown in (a) for the first sheet and in (b) for the second sheet of the dispersion curve.
Contour lines are labelled following the normalization $\beta p$} 
\label{fig03}
\end{figure}

In Fig.~\ref{fig03} we map different contours of isofrequency $\beta$ in the $k_x K$ plane based on the graphical representation of the dispersion equation given in Fig.~\ref{fig02}.
Thus we consider real values of $K$ again.
The first band is represented in Fig.~\ref{fig03}(a) providing a surface of maxima at $k_x = 0$ and $K = \pm \pi / p$ reaching $\beta_{max}$.
Simple closed curves around these points are found for decreasing values of the propagation constant provided $\beta_{c 1} \le \beta < \beta_{max}$, where $\beta_{c 1} p = 8.503$ coincides with the upper boundary of the first bandgap.
At lower $\beta$, spatial frequencies around $k_x = 0$ cannot excite propagating Bloch modes leading to open (isofrequency) curves.
This gap is removed in the second sheet of the dispersion equation, which arises for $\beta \le \beta_{c 2}$ and it is shown in Fig.~\ref{fig03}(b); obviously $\beta_{c 2} p = 8.301$ corresponds to the upper limit of the second band of allowed spatial frequencies $k_\parallel$.
In this case, the closed curves are centered at the origin of the plane $k_x K$ (and at multiples of $2 \pi / p$ along the $K$-axis).
Similarly to Fig.~\ref{fig03}(a), the isofrequency curves are closed when $\beta$ takes values belonging to the (second) band of allowed frequencies, here $\beta_{c 3} \le \beta < \beta_{c 2}$ where $\beta_{c 3} p = 7.899$.
Out of this interval, the isofrequency curves are open again.

\subsection{Bloch components of the wavefield}

According to Eq.~(\ref{eq14}), once the amplitudes $\mathbf{v}_0$ and $\mathbf{v}_1$ are given for all Bloch wavenumbers $K$, the periodic function $h_K$ and thus the wavefield $h_x$ are estimated unambiguously.
In fact we have shown that $\mathbf{v}_1 = \mathbf{P}_1^{-1} \mathbf{D}_1^{-1} \mathbf{D}_0 \mathbf{v}_0$, therefore the degrees of freedom are reduced to the amplitudes $\mathbf{v}_0$.
With the normalization $h_K (0) = 1$ we additionally impose that the summation of the elements of $\mathbf{v}_0$ is left to unity.
As a result, the amplitudes associated with the Bloch modes in the dielectric slab at $R_0 \equiv (-p + w,0]$ yield
\begin{equation}
 \mathbf{v}_0
 =
 \frac{1}{\exp \left( i K p \right) + B_0 - A_0}
 \left[
  \begin{array}{c}
   B_0 \\
   \exp \left( i K p \right) - A_0
  \end{array}
 \right] ,
\label{eq17}
\end{equation}
where $A_0$ and $B_0$ are matrix elements of $\mathbf{M}_0$.

We point out that the denominator in Eq.~(\ref{eq17}) vanishes if
\begin{equation}
 \frac{k_{y m}}{k_{y d}} = \frac{\left( - \epsilon_m \right)}{\epsilon_d} .
\label{eq16}
\end{equation}
which in principle might lead to a singular behavior of $\mathbf{v}_0$.
In this case we also have $B_0 = 0$. 
Eq.~(\ref{eq16}) represents the condition for the existence of a surface resonance in a metal-dielectric interface \cite{Raether88}.
Therefore $k_{y d}$ would be purely imaginary, which is associated with the existence of evanescent waves in the dielectric films.
In addition $\mathbf{M}_0$ would be a diagonal matrix with
\begin{equation}
 \pm K = \frac{w}{p} k_{y m} - \left( 1 - \frac{w}{p} \right) k_{y d} ,
\end{equation}
providing the eigenvector $\mathbf{v}_0$ as either the unit vector $[1, 0]$ or $[0, 1]$.
Moreover, $\mathrm{Im}(K) \neq 0$ except when $K$ vanishes, a case that occurs if additionally
\begin{equation}
 \frac{\left( - \epsilon_m \right)}{\epsilon_d} = \frac{p - w}{w} ,
\end{equation}
leading to $\mathbf{M}_0 = \mathbf{I}$.
The latter particular case might bring new insights onto our basic analysis however it deserves a thorough study out of here.

\begin{figure}
\centering
\includegraphics[width=8cm]{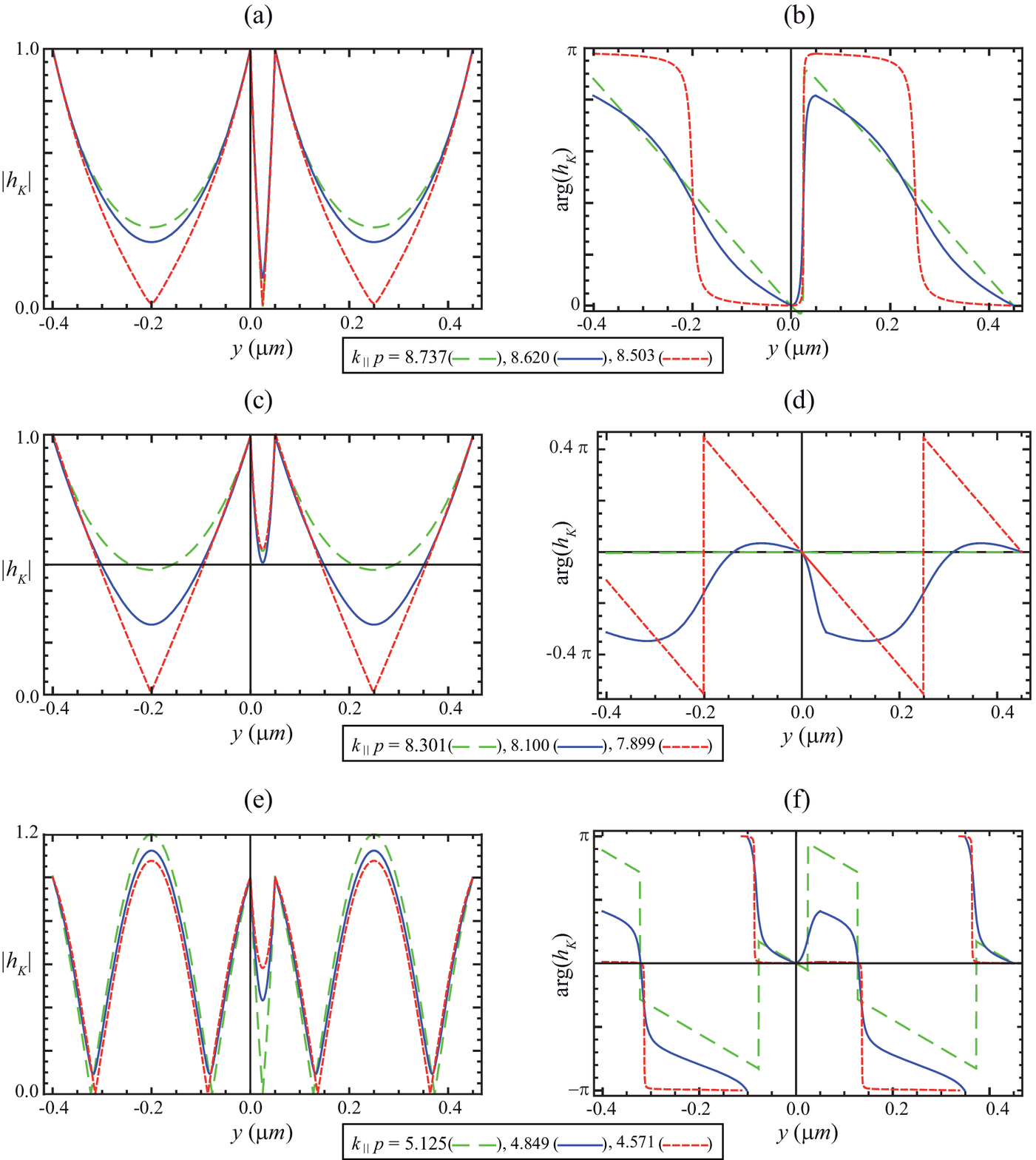}
\caption{Behavior of the complex amplitude (left column) and argument (right column) of the wavefunction $h_K$ at different values of $k_\parallel$ and positive $K$ corresponding to the first band (top), second band (middle) and third band (bottom).} 
\label{fig04}
\end{figure}

The periodic function $h_K$ is depicted in Fig.~\ref{fig04} for the multilayered setup with parameters given in the caption of Fig.~\ref{fig02} and at different values of $k_\parallel$.
In all cases, the complex amplitude $|h_K|$ is an even function whereas its argument is an odd function respect to the center of the metallic film, as shown in Eq.~(\ref{eq22}), provided it is set off to zero.
Peaks are formed at metal-dielectric interfaces and valleys at the center of the slabs; this rule holds in the purely-evanescent regime when $k_\parallel > \beta_c$, however the maximum of intensity is reached at the center of the dielectrics if $k_\parallel$ belongs to the third band [see Fig.~\ref{fig04}(e)]. 
On other hand we observe that phases of the wavefunction $h_K$ have a characteristic variation at different spectral bands.
For instance, phases are $\sim \pi\ \mathrm{rad}$ at the right side of the central metallic slab $y = w$ in the top band; however, this phase turns to $\sim 0$ in the second band.
We show below that this is of relevance in the formation of an on-axis focus.

\section{\label{sec03} Focus generation}

Let us establish some favourable conditions for the formation of a focus along the $z$-axis.
Using $(x,y) = (0,0)$ in Eq.~(\ref{eq05}) we obtain the wavefield amplitude $h_x = \sum_K \int a_K d k_x$ at the origin as a summation of the amplitudes $a_K$ corresponding to different Bloch modes. 
This may be interpreted as an interference of Bloch-type individuals.
If the phase of their amplitudes are manipulated in order to have the same value leading to in-phase waves, the oscillatory superposition yields the highest intensity achievable.
Excepting a few particular conditions, it cannot be found a point other than the origin from the $xy$ plane where such a phase matching holds.
As a consequence, a strong localization of the nondiffracting beam is expected to occur around the $z$-axis, such a line unquestionably constituting a focus.

At this point of our analysis it is interesting to review the concept of focus wavefields in a system such as the uniform dielectric medium, where nondiffracting solutions of the wave equation are well known.
Obviously we are speaking of Bessel beams \cite{Durnin87}.
Such a system would result from setting $w = 0$ in the layered medium of Fig.~\ref{fig01} to remove the metallic films from the dielectric host.
In this case Eq.~(\ref{eq09}) leads trivially to the solutions $K_\pm = \pm k_{y d}$.
Also $h_K (y) = 1$ in the whole $x y$ plane.
The phase matching condition at the origin is also observed over a radially-symmetric transverse pattern if in addition $a_K$ is in direct proportion to $|d \phi / d k_x| = |-1 / K|$, where $\phi$ is the polar angle in the $k_x K$ plane.
Thus inserting 
\begin{equation}
 a_{K_\pm} = \left\{ 
              \begin{array}{ll}
               \frac{1}{\sqrt{\beta_c^2 - \left( k_x^2 + \beta^2 \right)}} , & \mbox{ if } |k_x| < \sqrt{ \beta_c^2 - \beta^2 } \\
               0  ,                & \mbox{ otherwise}
              \end{array}
             \right.
\label{eq10}
\end{equation}
into Eq.~(\ref{eq05}) yields the Bessel wavefield $h_x = 2 \pi J_0 (k_\perp r)$ provided $\beta < \beta_c$ [see Eq.~(\ref{eq18})], being 
\begin{equation}
 k_\perp = \sqrt{\beta_c^2 - \beta^2}
\end{equation}
the transverse wave number and $r$ the radial spatial coordinate.
See Ref.~\cite{Zapata10a} for further details.
We point out that excitations of the type $a_K = 1 / |K|$  [if $\mathrm{Im} \left( K \right) = 0$, $a_K = 0$ otherwise] also provide paraxial Bessel beams in periodic media \cite{Miret08} and in anisotropic crystals \cite{Ciattoni03}.

A field spectrum like that of Eq.~(\ref{eq10}) may be experimentally attained using an opaque screen, with a centered extremely-thin transparent annulus, placed at the front focal plane of a perfect lens \cite{Indebetouw89}. 
Highly-efficient approaches may be found using conical lenses \cite{Grunwald00,Zapata06e} and mirrors \cite{Kuntz09}, Fabry–-Perot interferometers \cite{Horvath97}, leaky screens \cite{Holm98,Reivelt02b,Reivelt02}, and diffractive optical elements \cite{Vasara89,Amako03,Li09}.
Using such devices as external sources in our system would excite the required diffraction-free wavefields in the layered medium.
For simplicity we assume that the spectral strength of such nondiffracting beams has a form following Eq.~(\ref{eq10}),
\begin{equation}
 a_K = \frac{1}{\sqrt{\gamma^2 - k_x^2}} \mbox{, for } |k_x| < \gamma ,
\label{eq19}
\end{equation}
and $a_K = 0$ if $|k_x| \ge \gamma$.

Let us consider that $\beta$ and $\gamma$ may be tuned at convenience.
Thus $\beta \le k_\parallel < \sqrt{\beta^2 + \gamma^2}$ so that we might apply any spectral band (i.e. ordinates in Fig.~\ref{fig02}) arbitrarily.
In our model it is reasonable to think of $\sqrt{\beta^2 + \gamma^2}$ denoting the cut-off frequency $\bar{\beta}_c = 2 \pi \sqrt{\bar{\epsilon}_d} / \lambda_0$ associated with the uniform external medium of dielectric constant $\bar{\epsilon}_d$ where the source field is driven.
In a great number of examples given below $\bar{\beta}_c = \beta_{max}$ yielding $\bar{\epsilon}_d = 2.89$; in this medium the wavelength $\bar{\lambda}_d = \lambda_0 / \sqrt{\bar{\epsilon}_d} =  324\ \mathrm{nm}$ and the minimum spot size of the zero-order Bessel beam is $\left( \Delta_x \right)_{min} = 2.253 / \gamma = 116\ \mathrm{nm}$ where the maximum value of $\gamma = \bar{\beta}_c$ at $\beta = 0$.
In other hand, our model neglects filtering (apodizing) effects and aberrations on the wave fields induced at the boundaries of the system \cite{Zapata10a}.
However, conventional techniques for its compensation might be employed in order to find a good agreement with our results.
Otherwise the theory remains valid leaving the appropriate estimation of the spectrum $a_K$.

\section{\label{sec04} Subwavelength transverse patterns: Numerical results}

\begin{figure}
\centering
\includegraphics[width=7.5cm]{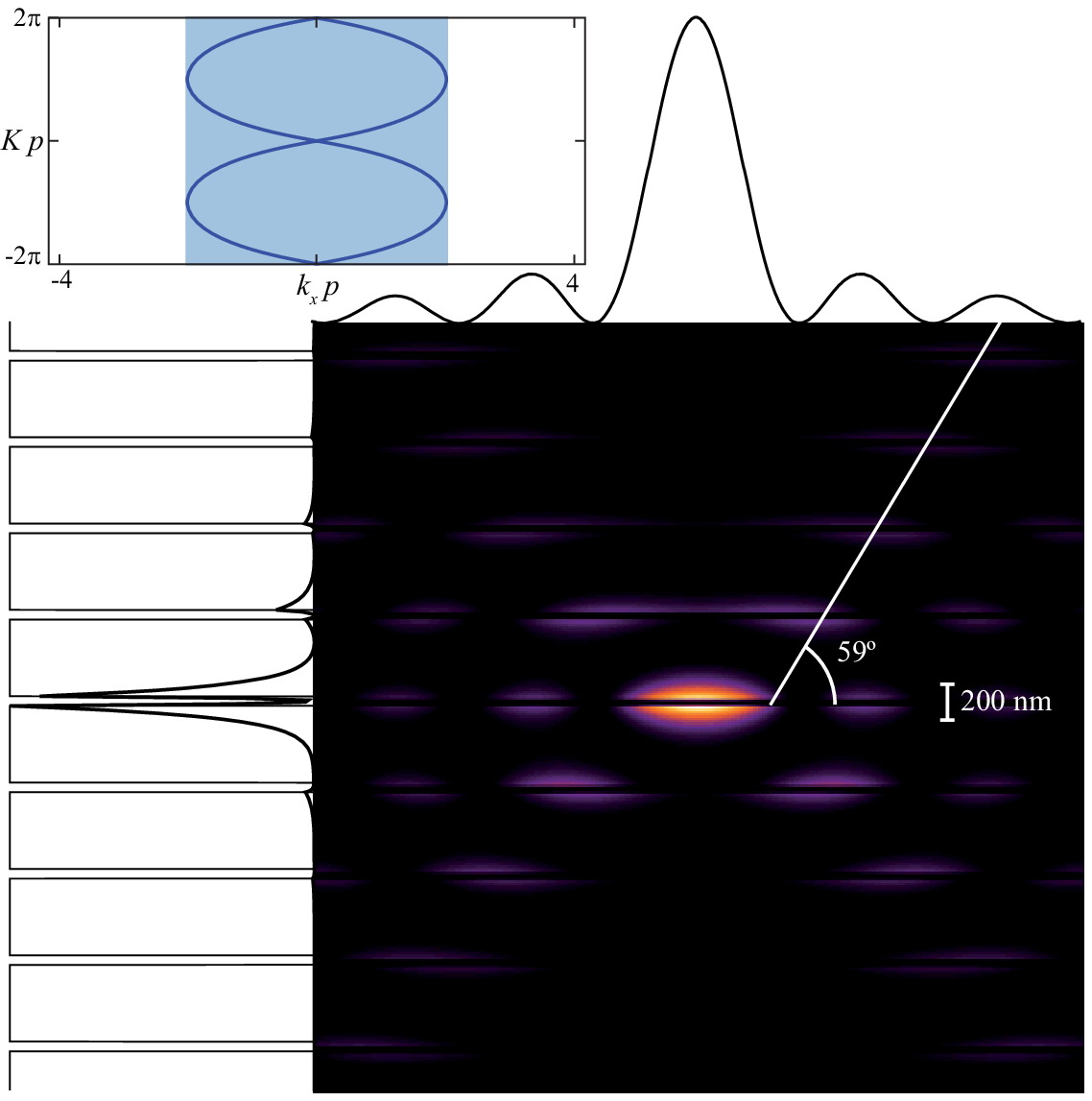}
\caption{Contour plot of the intensity $|h_x|^2$ in the $x y$ plane corresponding to a localized diffraction-free beam of normalized propagation constant $\beta p = 8.503$ and transverse frequency $\gamma p = 2.008$ leading to $\left( k_\parallel p \right)_{max} = 8.737$.
Intensity distributions along the coordinate axes are shown at the top and left sides.
Inset: Isofrequency curve where shaded region corresponds to the excited spatial bandwidth.} 
\label{fig05}
\end{figure}

To illustrate the focus generation along the $z$-axis, we perform a numerical simulation in the periodic layered medium of $p = 450\ \mathrm{nm}$ (and $\epsilon_d = 2.25$) with metallic films of $w = 50\ \mathrm{nm}$ and $\epsilon_m = - 15.0$.
In Fig.~\ref{fig05} we show the field intensity $|h_x|^2$ for a nondiffracting beam of propagation constant $\beta = 18.90\ \mu\mathrm{m}^{-1}$ ($\beta = \beta_{c 1}$), which spatial spectrum is given by Eq.~(\ref{eq19}) with $\gamma = 4.462\ \mu\mathrm{m}^{-1}$ ($\gamma p = 2.008$).
Thus the maximum value of $k_\parallel$ reaches a value coinciding with $\beta_{max}$, i.e. $\beta_{c1} \le k_\parallel < \beta_{max}$ corresponding to the first band of allowed frequencies shown in Fig.~\ref{fig02}.
The isofrequency curve $K = K(k_x)$ is also depicted in the inset of the figure, where excited frequencies $k_x$ are shaded in blue.
The intensity is maximum at the origin but the pattern exhibits no radial symmetry.
The field distribution along the abscissa, 
\begin{equation}
 h_x (x,y = 0) = 2 \int_{-\pi/2}^{\pi/2} \cos \left( \gamma x \cos \phi \right) d \phi = 2 \pi J_0 (\gamma x) , 
\label{eq21}
\end{equation}
resembles that of the source field that would propagate in the homogeneous dielectric medium of 
$\bar{\epsilon}_d = 2.89$.
We point out that some discrepancies arise in Eq.~(\ref{eq21}) if some spatial frequencies $k_x$ are excited but, because of the presence of a gap, they do not contribute to the resulting propagating wavefield; this case will be treated ahead.
From Eq.~(\ref{eq21}) we derive that the FWHM of the intensity peak along this direction is inversely proportional to $\gamma$ following $\Delta_x = 2.253 / \gamma$.
In our case $\Delta_x = 505\ \mathrm{nm}$, which is above the wavelength in the dielectric host $\lambda_d = \lambda_0 / \sqrt{\epsilon_d}$ ($ = 367\ \mathrm{nm}$).
However, the behavior along the ordinate is significantly different.
The most attractive feature of the wavefield is its high localization in the metal-dielectric interfaces, leading to fast decays when moving away from the surfaces and thus forming wedge-like shapes.
This is in agreement with the patterns shown in Figs.~\ref{fig04}(a) and (b) contributing to the integral (\ref{eq05}) at $x = 0$.
Although the highest peak is attained at $y = 0$, a large one also arises on the other side of the central metallic film, $y = w$.
There, the Bloch modes $h_K (y) \exp \left( i K y \right)$ are as strong as in focus and they interfere nearly in-phase (dephase $ < 0.29 \pi\ \mathrm{rad}$) giving a secondary focus.
Ignoring this sidelobe, the FWHM of the figure is $\Delta_y = 45.99\ \mathrm{nm}$, well below $\lambda_d$. 

Apart from surface resonances, radial asymmetry from focus is also attributed to a high concentration of light along certain directions in the $xy$ plane.
This directional enhancement of the radiated power may be explained by means of an effect coined as photon focusing \cite{Chigrin04}, a term which first was used onto the strong anisotropy of heat flux in crystalline solids and that takes other names in optics like self-collimation \cite{Kosaka99}, self-guiding \cite{Chigrin03}, and subdiffractive propagation \cite{Staliunas06b}.

The estimation of the value(s) of the azimuthal coordinate $\theta$ where photon focusing is manifested is based on the stationary-phase principle.
Firstly note that the phase factor $\exp \left( i k_x x + i K y \right)$ of Eq.~(\ref{eq05}) varies rapidly when moving away from the origin, $r \to \infty$, being $(x,y) = r (\cos \theta,\sin \theta)$.
This phase term also introduces strong oscillations for running $k_x$ except in the vicinities of those spatial frequencies $k_{x s}$ satisfying $x + K'_s y = 0$, where $K' = d K / d k_x$ and the subindex $s$ stands for the value given at $k_{x s}$.
Setting $K'_s = \tan \phi_s$ we have the solutions $\phi_s = \theta \pm \pi / 2$; consequently the tangent of the curve $K(k_x)$ at the stationary points $k_{x s}$ is normal to the vector position of the observation point $(x,y)$.
Moreover, the asymptotic behavior of Eq.~(\ref{eq05}) depends exclusively on those stationary points $k_{x s}$; substituting $K$ by $K_s$ in every term of the integrand excepting the phase factor for which we use a quadratic expansion $K \approx K_s + K'_s (k_x - k_{x s}) + K''_s (k_x - k_{x s})^2 / 2$, finally yields
\begin{equation}
 h_x (x,y) \approx \sum_{K,k_{x s}} a_{K_s} h_{K_s} (y) \exp \left( i k_{x s} x + i K_s y \right) I_{K_s} (y) ,
\end{equation}
where
\begin{eqnarray}
 I_{K_s} (y) &=& \int_{-\infty}^{\infty} \exp \left[ i K''_s (k_x - k_{x s})^2 y / 2 \right] d k_x \nonumber \\
             &=& \sqrt{\frac{2 \pi}{|K''_s y|}} \exp \left[ i \frac{\pi}{4} \mathrm{sign} \left( K''_s y \right) \right] .
\end{eqnarray}
In this simple analysis we have also assumed that $a_K$ has a well behavior and it does not present discontinuities.
In the far field zone $|h_x|^2$ decreases inversely proportional to the radial coordinate $r$, excepting those directions $\theta$ where stationary points satisfy $K''_s = 0$.
In these cases the dispersion curve is flat leading to a significantly slow attenuation of the radiated power, at least much slower than $r^{-1}$.

\begin{figure}
\centering
\includegraphics[width=8cm]{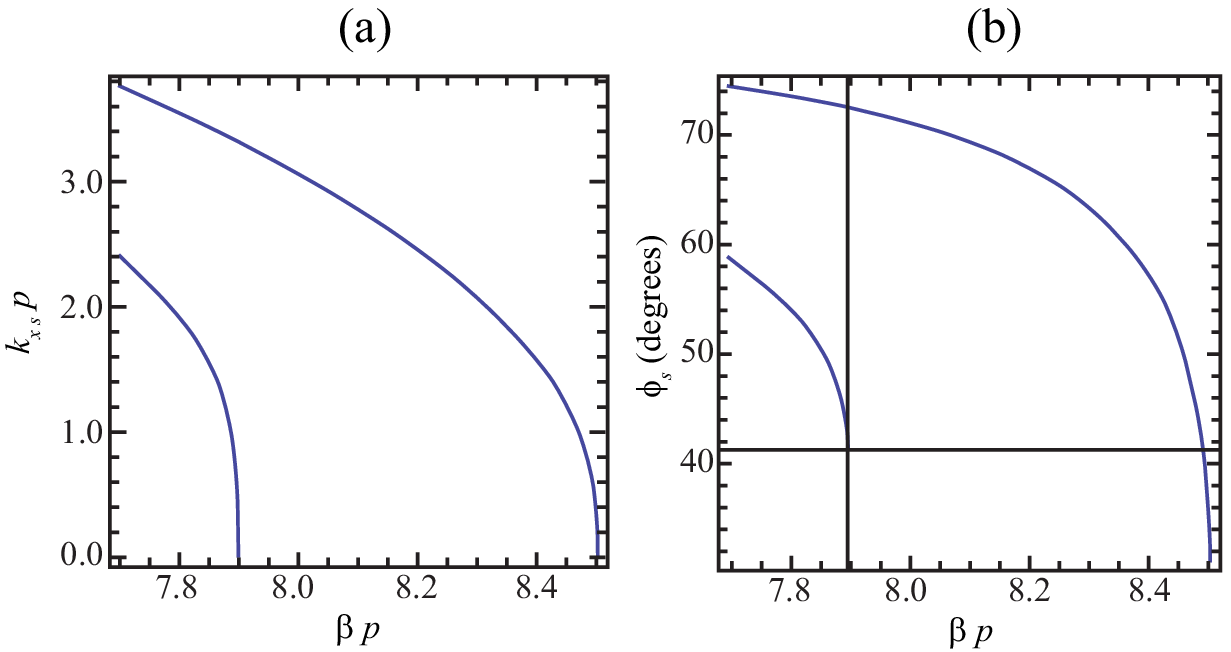}
\caption{(a) Solutions $k_x$ of the equation (\ref{eq20}), $K''_s = 0$, associated with different propagation constant $\beta$ of the wavefields, in the layered structure of Fig.~\ref{fig01}.
(b) Azimuthal angle $\phi$ in the plane $k_x K$ that corresponds to each solution of the aforementioned self-guiding condition.} 
\label{fig06}
\end{figure}

In Fig.~\ref{fig06}(a) we plot the solutions $k_{x s}$ of the equation $K''_s = 0$ at different values of $\beta$.
Considering that the dispersion equation (\ref{eq09}) may be written as $\cos (K p) = f(k_x)$, the solutions of the equation $K''_s = 0$ are also the frequencies $k_{x s}$ satisfying
\begin{equation}
 f''_s \left( 1 - f_s^2 \right) + f'^2_s f_s = 0 .
\label{eq20}
\end{equation}
As shown in Fig.~\ref{fig03}(a), the isofrequency curves for $\beta_{c 1} < \beta < \beta_{max}$ mimic ellipses that lack of flat sections so that $K''_s = 0$ has no real solutions.
At $\beta = \beta_{c 1}$, the isofrequency curve represented also in the inset of Fig.~\ref{fig05} is flat at the origin of the plane $k_x K$ taking a shape in X.
In this case $(k_{x s},K_s) = (0,0)$ for which $K'_s = \pm 0.604$ and $\phi_s = \pm 31\ \mathrm{deg}$.
See Fig.~\ref{fig06}(b) showing the estimated (positive) values of $\phi_s$ also for other propagation constants, $\beta$.
Our analysis conclude that the self-guiding phenomenon is expected at angles $\theta = \pm 59\ \mathrm{deg}$ (and $\theta = \pm 121\ \mathrm{deg}$), in agreement with the transverse pattern shown in Fig.~\ref{fig05}.
When $\beta_{c 3} < \beta < \beta_{c 1}$, solving Eq.~(\ref{eq20}) leads to increasing values of $|k_{x s}|$ and therefore increasing values of $|\phi_s|$, as shown in Figs.~\ref{fig06}(a) and (b) respectively.
Interestingly, $k_{x s} = 0$ becomes again a solution if $\beta = \beta_{c 3}$ (here giving $K_s = \pm \pi / p$) for which $\phi_s = \pm 42\ \mathrm{deg}$; also $k_{x s} = \pm 3.32 / p$ gives $K''_s = 0$ where $\phi_s = \pm 73\ \mathrm{deg}$.
Moreover, a minimum of four values of $k_{x s}$ may be found for $\beta < \beta_{c 3}$ in which the zero-diffraction condition $K''_s = 0$ holds.
We may conclude the general rule that a larger number of solutions comes out when (positive) $\beta$ decreases reaching new bandgaps for the frequency $k_\parallel$ as shown in Fig.~\ref{fig02}.

\begin{widetext}

\begin{figure}
\centering
\includegraphics[width=16cm]{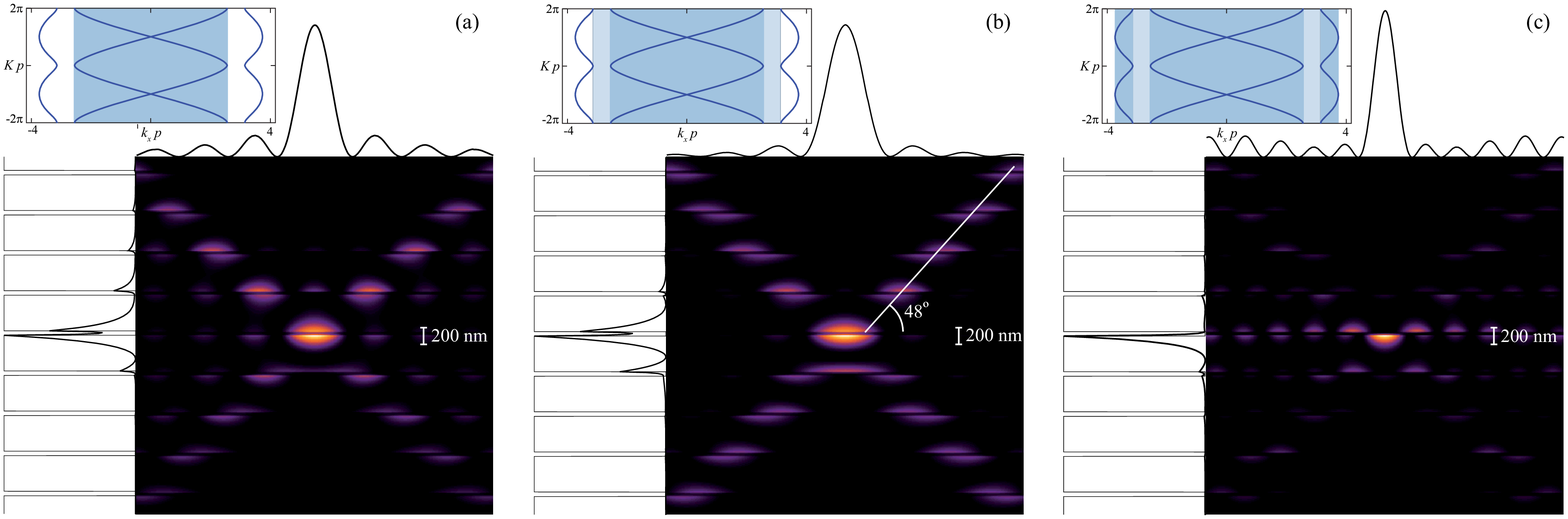}
\caption{Transverse intensity $|h_x|^2 (x,y)$ for diffraction-free beams of $\beta p = 7.899$ and spatial frequency (a) $\gamma p = 2.552$, (b) $\gamma p = 3.148$, and (c) $\gamma p = 3.734$.
Insets: Isofrequency curve at $\beta p = 7.899$.
The blue shade indicates the spectral window in $k_x$ under excitation.
Light blue regions refer to gap-induced frustrated excitation.} 
\label{fig07}
\end{figure}

\end{widetext}

Let us analyze the intensity patterns of diffraction-free beams under the presence of bandgaps [$\mathrm{Im} (K) \neq 0$] within the source spectral window $|k_x| < \gamma$.
In Fig.~\ref{fig07} we plot $|h_x|^2$ of wavefields with propagation constant $\beta = \beta_{c 3}$ corresponding to the lower limit of the second band.
In Fig.~\ref{fig07}(a) $\gamma p = 2.552$ that yields $\left( k_\parallel \right)_{max} = \beta_{c 2}$.
This means that the source excites entirely the second band of allowed spatial frequencies with no gaps; therefore Eq.~(\ref{eq21}) is valid and $\Delta_x = 2.253 / \gamma$ ($= 397\ \mathrm{nm}$).
Along the $y$-axis, the narrow peak around the origin has a FWHM $\Delta_y = 62.52\ \mathrm{nm}$ and again is accompanied by a high sidelobe at the other side of the metallic film.
The formation of this secondary focus is again the constructive interference (nearly in-phase) of the Bloch modes from the second band.
On other hand, self-collimation is attributed exclusively to the stationary points $(k_{x s},K_s) = (0,\pm \pi / p)$ giving $\theta = \pm 48\ \mathrm{deg}$ ($\phi_s = \pm 42\ \mathrm{deg}$ as seen above).

Setting $\gamma p = 3.148$ as shown in Fig.~\ref{fig07}(b), the bandgap $\beta_{c 2} < k_\parallel < \beta_{c 1}$ is encountered, that is frequencies satisfying $\gamma_{c 1} < |k_x| < \gamma_{c 2}$ are frustratedly excited (being $\gamma_{c 1} p = 2.552$ and $\gamma_{c 2} p = 3.148$), so that the optical system behaves like a low-pass filter.
In this case
\begin{equation}
 h_x (x,y = 0) = 2 \pi J_0 (\gamma x) - h_{x c 1} (\gamma x), 
\end{equation}
where
\begin{equation}
 h_{x {c 1}} (\gamma x) = 2 \int_{-\phi_{c 1}}^{\phi_{c 1}} \cos \left( \gamma x \cos \phi \right) d \phi = \sum_{m = 0}^\infty C_m^{({c 1})} J_{2 m} \left( \gamma x \right) 
\end{equation}
being $0 \le \phi_{c 1} < \pi / 2$ such that $\gamma \cos \phi_{c 1} = \gamma_{c 1}$, 
\begin{eqnarray}
 C_m^{({c 1})} = 4 \left( -1 \right)^{m} \frac{\sin \left( 2 m \phi_{c 1} \right)}{m} \mbox{, for $m \neq 0$ ,}
\end{eqnarray}
and $C_0^{({c 1})} = 4 \phi_{c 1}$.
We point out that $J_{2 m} \left( 0 \right) = 0$ for $m \neq 0$ so that in the vicinities of the focal point we have $h_x \approx \left( 2 \pi - 4 \phi_{c 1} \right) J_0 (\gamma x)$.
However the central peak stretches slightly, specifically $\Delta_x = 459\ \mathrm{nm}$, whereas sidelobes are altered significantly as expected.
Moreover, FWHM in the direction of the periodicity $\Delta_y = 61.58\ \mathrm{nm}$ and self-collimation angles $\theta = \pm 48\ \mathrm{deg}$ are also maintained is spite of bandgaps.

Finally, arriving at $\gamma p = 3.734$ we are able to excite every allowed spatial frequency in the first band since $\left( k_\parallel \right)_{max} = \beta_{max}$.
The intensity $|h_x|^2$ is depicted in Fig.~\ref{fig07}(c) showing notable differences from the cases previously analyzed.
For instance, the FWHM along the $x$-axis is $\Delta_x = 277\ \mathrm{nm}$, which is close to the value we would obtain if we ignore the bandgap.
In this case $\Delta_x$ is lower than the radiation wavelength in the dielectric $\lambda_d = 367\ \mathrm{nm}$; however this subwavelength size is still higher than $\lambda_d / 2$ in the same order than a regular Bessel beam.
Here we may evaluate the wavefield by
\begin{equation}
 h_x (x,y = 0) = 2 \pi J_0 (\gamma x) - h_{x {c 1}} (\gamma x) + h_{x {c 2}} (\gamma x), 
\end{equation}
where $\gamma \cos \phi_{c 2} = \gamma_{c 2}$.
Sidelobes are also strongly attenuated in all directions; even at those associated with self-collimation.
Here $\theta = \pm 48\ \mathrm{deg}$ ($\phi_s = \pm 42\ \mathrm{deg}$) and also $\theta = \pm 17\ \mathrm{deg}$ ($\phi_s = \pm 73\ \mathrm{deg}$); as a consequence, sharing the power radiated in this large number of directions leads to the weakening of the photon-focusing effect.
More importantly, the high sidelobe appearing previously in the $y$-axis seems to be wiped out completely.
This effect may be explained considering that Bloch components from the first band and those from the second band are roughly out of phase at $y = w$ [see Figs.~\ref{fig04}(b) and (d)] so that they interfere destructively in Eq.~(\ref{eq05}).
Additionally, the central peak is unaltered in practical terms, giving $\Delta_y = 54\ \mathrm{nm}$.

\section{\label{sec05} Hybrid construction of diffraction-free waves}

In the numerical simulations given above we have shown that beamsizes along the $x$-axis are larger than the diffraction limit $\lambda_d / 2$ attained by quasi-stationary Bessel beams propagating in the medium of dielectric constant $\epsilon_d$, whereas $\Delta_y$ is clearly subwavelength.
Control over the wave pattern and thus over its FWHM in the $x$ direction is exercised by the spectrum of spatial frequencies $k_x$: the higher bandwidth the lower $\Delta_x$.
From $k_x^2 \le \beta_{max}^2 - \beta^2$ estimating the spectral domain of the wavefield we conclude that decreasing $\beta$ leads to a widening of the spectrum.
However, if $\beta < \beta_c$ the plane-wave components in the dielectric material propagate homogeneously so that they are not necessarily coupled resonantly around the metal-dielectric surfaces.
In principle, this effect might modify significantly the localization of the field in the direction of the periodicity.
Let us clarify these aspects.

\begin{figure}[t]
\centering
\includegraphics[width=7.5cm]{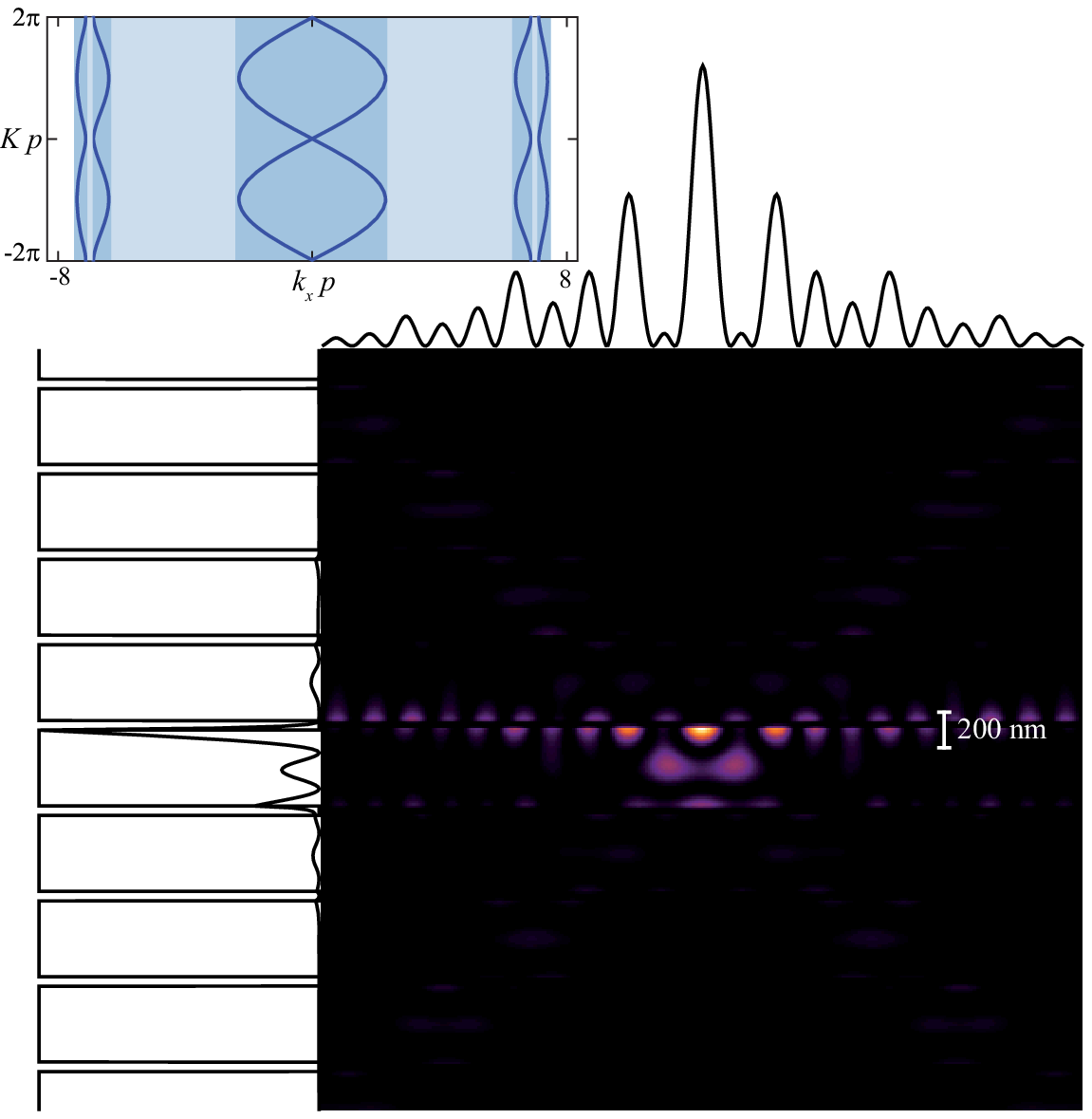}
\caption{Intensity $|h_x|^2$ in the transverse $xy$ plane for a nondiffracting beam of $\beta = 10.16\ \mu\mathrm{m}^{-1}$ and $\gamma = 16.55\ \mu\mathrm{m}^{-1}$. 
Excitation of Bloch modes with $|k_x| <  5.15\ \mu\mathrm{m}^{-1}$ leads to nonevanescent wavelets in the dielectric slabs, so that the resulting hybrid wavefield from Eq.~(\ref{eq05}) combines Bloch constituents of different nature.}
\label{fig08}
\end{figure}

The contour plot shown in Fig.~\ref{fig08} corresponds to a diffraction-free beam of propagation constant $\beta = 10.157\ \mu\mathrm{m}^{-1}$ ($\beta p = 4.571$) coinciding with the lower limit of $k_\parallel$ for the third band (see Fig.~\ref{fig02}).
The transverse frequency $\gamma p = 7.446$ guarantees that all the three bands of allowed frequencies are excited; this is a high value since if it were generated in a uniform medium with $\bar{\epsilon}_d = 2.89$ would give a Bessel beam of width $\Delta_x = 136\ \mathrm{nm}$.
Fig.~\ref{fig08} shows that the intensity $|h_x|^2$ is distributed mainly around the interface $y = 0$.
A narrow peak on the focus is produced exhibiting a width $\Delta_x = 132\ \mathrm{nm}$, which is slightly lower than that just given above.
Also strong sidelobes arise in the vicinity of focus.
This suggests that bandgaps inherent in a photonic crystal provide a mechanism to achieve a superresolving effect going beyond the diffraction-induced Bessel limit.
The origin of this superresolving effect is not the excitation of SPPs but an enhancement of high spatial frequencies \cite{Martinez99}.

On other hand, the field distribution in the $y$-axis demonstrates a subwavelength focus of FWHM $\Delta_y = 44\ \mathrm{nm}$.
Sidelobes on the interfaces are accompanied with others peaks in the middle of the dielectric slabs.
This is not surprising since Bloch components of the third band shown in Fig.~\ref{fig04}(e) and contributing in the expansion (\ref{eq05}) have such a behavior.
This effect is in fact more pronounced along directions oriented with polar angles $\theta = \phi_s \pm \pi/2$ being the stationary point $\phi_s = \pm 41\ \mathrm{deg}$ ($\theta = \pm 49\ \mathrm{deg}$) associated with the self-collimation effect.

\begin{figure}[t]
\centering
\includegraphics[width=7.5cm]{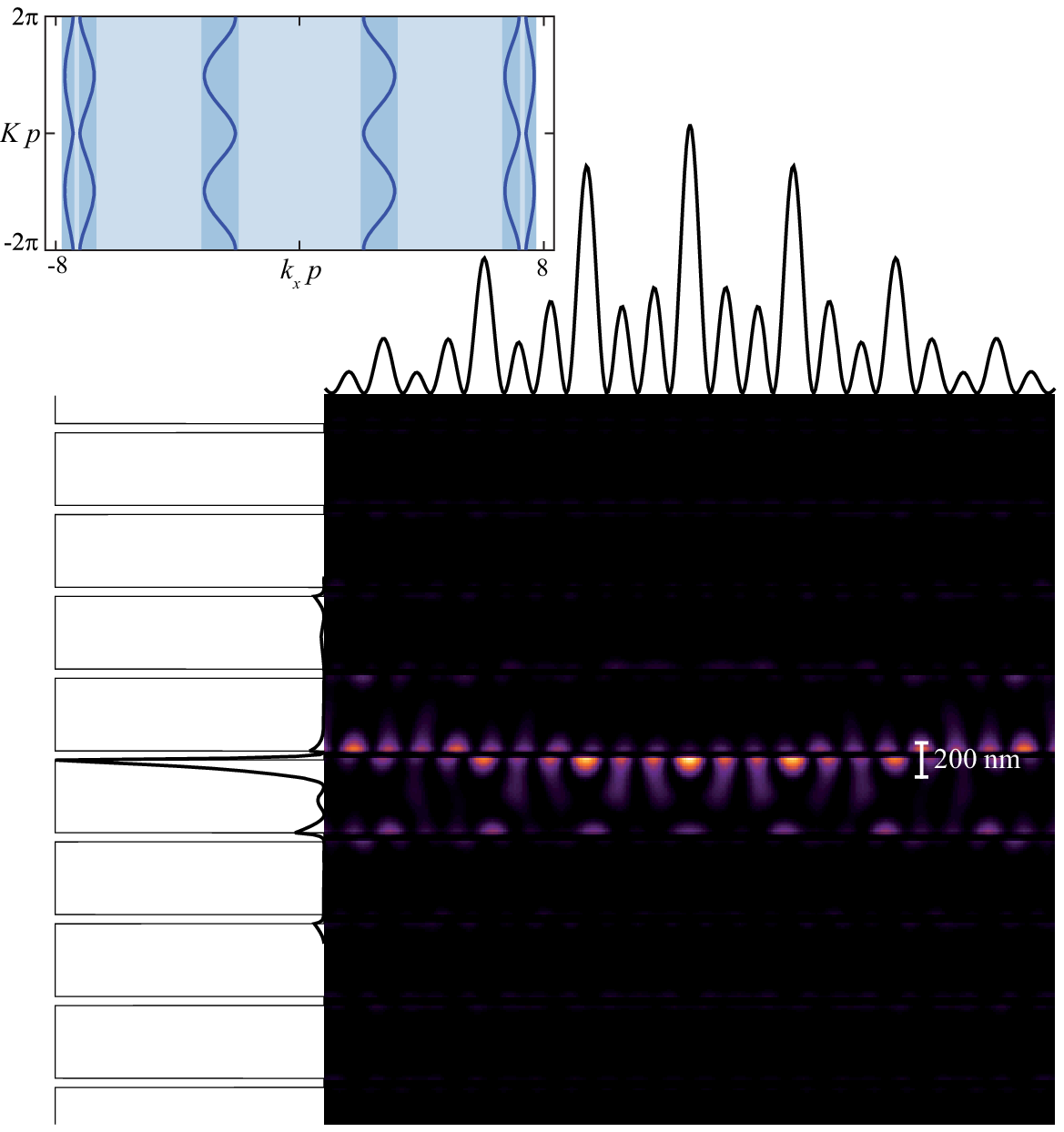}
\caption{ Transverse pattern of a nondiffracting beam with $\beta = 9.00\ \mu\mathrm{m}^{-1}$.
The beamwidth of the main central focus has $\Delta_x = 109\ \mathrm{nm}$ and $\Delta_y = 46\ \mathrm{nm}$, going beyond the diffraction limit $\Delta_{min} = 116\ \mathrm{nm}$.} 
\label{fig09}
\end{figure}

Although superresolution along the $x$-axis is modest in comparison with that obtained in the $y$-axis, we wonder if beamwidths might surpass the diffraction limit in all directions.
As resolved in Sec.~\ref{sec03}, for our examples where $\beta_{max} = 19.42\ \mu\mathrm{m}^{-1}$ and $\lambda_0 = 550\ \mathrm{nm}$ this diffraction limit reads $\Delta_{min} = 116\ \mathrm{nm}$.
In Fig.~\ref{fig09} we represent the intensity distribution $|h_x|^2$ of a nondiffracting beam excited with a transverse spatial frequency $\gamma = 17.20\ \mu\mathrm{m}^{-1}$ giving a propagation constant $\beta = 9.00\ \mu\mathrm{m}^{-1}$.
In this case the central peak is anamorphic and characterized by FWHMs $\Delta_x = 109\ \mathrm{nm}$ and $\Delta_y = 46\ \mathrm{nm}$ all below the diffraction limit.
The bandgap-induced filtering also leads to intense sidelobes along the interface $y = 0$, however they decay as going beyond the focus.
Moreover, the peak intensity of these secondary foci cannot exceed that intensity of the main focus.

\section{\label{sec06} Concluding remarks}

We have demonstrated that diffraction-free beams propagating in structured media composed of alternating layers of positive and negative $\epsilon$ may reach beamsizes surpassing the diffraction limit.
For an excitation of the form given in (\ref{eq19}), we recall that this limit is attained when the transverse spatial frequency is maximum, $\gamma_{max} = \sqrt{\bar{\epsilon}_d} \omega / c$, leading to stationary ($\beta = 0$) Bessel waves with a central spot of FWHM $\Delta_{min} = 2.253 / \gamma_{max}$.
In the layered system, two different mechanisms lead to the superresolving effect.
A bandpass filtering due to the existence of gaps in the spatial spectrum of $k_x$ modifies the response of the system transversally providing a narrow peak along the $x$-axis with moderate gain ($\Delta_x \lesssim \Delta_{min}$) and high sidelobes (secondary foci).
In the direction of the periodicity, however, superresolution is carried out by the formation of surface resonances in the metal-dielectric interfaces leading to fast decays out of these planes.
In our numerical simulations, $\Delta_y$ can be as low as a third of $\Delta_{min}$.
Additionally, dephase of Bloch constituents belonging to different bands contributes to the growth control of secondary foci in nearby surfaces.

In principle, subwavelength beamwidths along an infinitely long distance might potentially be obtained.
However practical limitations in the geometry of the system leads to the generation of invariant wavefields along with a finite axial distance.
The estimation of such limits is similar for example to that corresponding to Bessel beams propagating in free space \cite{Durnin87,Durnin87b}, and therefore they are not discussed here.
However we point out the decisive role of the structured medium in order to achieve the subwavelength beamsizes: these wavefields cannot be sustained in free space for ranges longer than a wavelength.
Relevant aspects in near-field subwavelength nondiffracting beams are exhaustively investigated by Kukhlevsky and co-workers \cite{Kukhlevsky03,Kukhlevsky04,Kukhlevsky04b,Nyitray08}.

Finally, material losses are neglected in our analysis.
Since $\mathrm{Im}(\epsilon) \neq 0$ leads to wave damping in an unbounded dissipative system, absorption in the medium should be accounted for assuming a limited number of slabs.
We have performed numerical simulations showing that the superresolving effect along the $y$-axis is practically maintained for a small absorption coefficient of the metallic medium and a moderate number of layers, whereas the peak slightly gets wider along the $x$-axis.
Even assuming that the focus is placed at an interface near the center of the system, boundary effects and also tunnelling effects cannot be ignored.
For instance, setting $\epsilon_m = -15 + i 0.15$ we may excite a diffraction-free beam with a propagation constant $\beta = 9\ \mu\mathrm{m}^{-1}$ and a central peak of $\Delta_x = 139\ \mathrm{nm}$ and $\Delta_y = 43\ \mathrm{nm}$ in a medium composed of $20$ unit cells.
The deterioration in resolution power along the $x$-axis is mainly caused by a selective increment in $\mathrm{Im} (K)$ within the allowed spectral bands, an effect that is specially dramatic in the highest band.
We have estimated that the central peak is also enlarged to a lesser extent by increasing the number of layers.
Under these unfavourable conditions the beamsize is still surpassing the diffraction limit along the direction of the periodicity.
On other hand, a significant attenuation of the focal intensity is also evident.
In our numerical simulation, in-focus intensity decreases down to $18 \%$ (down to $6 \%$) from its original value if $\mathrm{Im} \left( \epsilon_m \right) = 0.15$ and the number of unit cells is $10$ ($20$).
A manuscript containing an extended analysis with further details is in preparation.

\begin{acknowledgments}                            
This research was funded by Ministerio de Ciencia e Innovaci\'on (MICIIN) under the project TEC2009-11635.
\end{acknowledgments}

\listoffigures
\printfigures



\begin{thebibliography}{10}%
\makeatletter
\providecommand \@ifxundefined [1]{%
 \ifx #1\undefined \expandafter \@firstoftwo
 \else \expandafter \@secondoftwo
\fi
}%
\providecommand \@ifnum [1]{%
 \ifnum #1\expandafter \@firstoftwo
 \else \expandafter \@secondoftwo
\fi
}%
\providecommand \enquote [1]{``#1''}%
\providecommand \bibnamefont  [1]{#1}%
\providecommand \bibfnamefont [1]{#1}%
\providecommand \citenamefont [1]{#1}%
\providecommand\href[0]{\@sanitize\@href}%
\providecommand\@href[1]{\endgroup\@@startlink{#1}\endgroup\@@href}%
\providecommand\@@href[1]{#1\@@endlink}%
\providecommand \@sanitize [0]{\begingroup\catcode`\&12\catcode`\#12\relax}%
\@ifxundefined \pdfoutput {\@firstoftwo}{%
 \@ifnum{\z@=\pdfoutput}{\@firstoftwo}{\@secondoftwo}%
}{%
 \providecommand\@@startlink[1]{\leavevmode}%
 \providecommand\@@endlink[0]{}%
}{%
 \providecommand\@@startlink[1]{%
  \leavevmode
  \pdfstartlink
   attr{/Border[0 0 1 ]/H/I/C[0 1 1]}%
   user{/Subtype/Link/A<</Type/Action/S/URI/URI(#1)>>}%
  \relax
 }%
 \providecommand\@@endlink[0]{\pdfendlink}%
}%
\providecommand \url  [0]{\begingroup\@sanitize \@url }%
\providecommand \@url [1]{\endgroup\@href {#1}{\urlprefix}}%
\providecommand \urlprefix [0]{URL }%
\providecommand \Eprint[0]{\href }%
\@ifxundefined \urlstyle {%
  \providecommand \doi [1]{doi:\discretionary{}{}{}#1}%
}{%
  \providecommand \doi [0]{doi:\discretionary{}{}{}\begingroup
  \urlstyle{rm}\Url }%
}%
\providecommand \doibase [0]{http://dx.doi.org/}%
\providecommand \Doi[1]{\href{\doibase#1}}%
\providecommand \bibAnnote [3]{%
  \BibitemShut{#1}%
  \begin{quotation}\noindent
    \textsc{Key:}\ #2\\\textsc{Annotation:}\ #3%
  \end{quotation}%
}%
\providecommand \bibAnnoteFile [2]{%
  \IfFileExists{#2}{\bibAnnote {#1} {#2} {\input{#2}}}{}%
}%
\providecommand \typeout [0]{\immediate \write \m@ne }%
\providecommand \selectlanguage [0]{\@gobble}%
\providecommand \bibinfo [0]{\@secondoftwo}%
\providecommand \bibfield [0]{\@secondoftwo}%
\providecommand \translation [1]{[#1]}%
\providecommand \BibitemOpen[0]{}%
\providecommand \bibitemStop [0]{}%
\providecommand \bibitemNoStop [0]{.\EOS\space}%
\providecommand \EOS [0]{\spacefactor3000\relax}%
\providecommand \BibitemShut [1]{\csname bibitem#1\endcsname}%
\bibitem{Durnin87}%
  \BibitemOpen
  \bibfield{author}{%
  \bibinfo {author} {\bibfnamefont{J.}~\bibnamefont{Durnin}},\ }%
  \bibfield{journal}{%
  \bibinfo {journal} {J. Opt. Soc. Am. A}\ }%
  \textbf{\bibinfo {volume} {4}},\ \bibinfo {pages} {651} (\bibinfo {year}
  {1987})%
  \bibAnnoteFile{NoStop}{Durnin87}%
\bibitem{Durnin87b}%
  \BibitemOpen
  \bibfield{author}{%
  \bibinfo {author} {\bibfnamefont{J.}~\bibnamefont{Durnin}}, \bibinfo {author}
  {\bibfnamefont{J.~J.}\ \bibnamefont{Miceli}},\ and\ \bibinfo {author}
  {\bibfnamefont{J.~H.}\ \bibnamefont{Eberly}},\ }%
  \bibfield{journal}{%
  \bibinfo {journal} {Phys. Rev. Lett.}\ }%
  \textbf{\bibinfo {volume} {58}},\ \bibinfo {pages} {1499} (\bibinfo {year}
  {1987})%
  \bibAnnoteFile{NoStop}{Durnin87b}%
\bibitem{Recami08}%
  \BibitemOpen
  \emph{\bibinfo {title} {Localized Waves}},\ edited by\ \bibinfo {editor}
  {\bibfnamefont{H.~E.}\ \bibnamefont{Hern\'andez-Figueroa}}, \bibinfo {editor}
  {\bibfnamefont{M.}~\bibnamefont{Zamboni-Rached}},\ and\ \bibinfo {editor}
  {\bibfnamefont{E.}~\bibnamefont{Recami}}\ (\bibinfo {publisher} {John Wiley
  \& Sons},\ \bibinfo {year} {2008})%
  \bibAnnoteFile{NoStop}{Recami08}%
\bibitem{Indebetouw89}%
  \BibitemOpen
  \bibfield{author}{%
  \bibinfo {author} {\bibfnamefont{G.}~\bibnamefont{Indebetouw}},\ }%
  \bibfield{journal}{%
  \bibinfo {journal} {J. Opt. Soc. Am. A}\ }%
  \textbf{\bibinfo {volume} {6}},\ \bibinfo {pages} {150–} (\bibinfo {year}
  {1989})%
  \bibAnnoteFile{NoStop}{Indebetouw89}%
\bibitem{Fagerholm96}%
  \BibitemOpen
  \bibfield{author}{%
  \bibinfo {author} {\bibfnamefont{J.}~\bibnamefont{Fagerholm}}, \bibinfo
  {author} {\bibfnamefont{A.~T.}\ \bibnamefont{Friberg}}, \bibinfo {author}
  {\bibfnamefont{J.}~\bibnamefont{Huttunen}}, \bibinfo {author}
  {\bibfnamefont{D.~P.}\ \bibnamefont{Morgan}},\ and\ \bibinfo {author}
  {\bibfnamefont{M.~M.}\ \bibnamefont{Salomaa}},\ }%
  \bibfield{journal}{%
  \bibinfo {journal} {Phys. Rev. E}\ }%
  \textbf{\bibinfo {volume} {54}},\ \bibinfo {pages} {4347} (\bibinfo {year}
  {1996})%
  \bibAnnoteFile{NoStop}{Fagerholm96}%
\bibitem{Mugnai00}%
  \BibitemOpen
  \bibfield{author}{%
  \bibinfo {author} {\bibfnamefont{D.}~\bibnamefont{Mugnai}}, \bibinfo {author}
  {\bibfnamefont{A.}~\bibnamefont{Ranfagni}},\ and\ \bibinfo {author}
  {\bibfnamefont{R.}~\bibnamefont{Ruggeri}},\ }%
  \bibfield{journal}{%
  \bibinfo {journal} {Phys. Rev. Lett.}\ }%
  \textbf{\bibinfo {volume} {84}},\ \bibinfo {pages} {4830} (\bibinfo {year}
  {2000})%
  \bibAnnoteFile{NoStop}{Mugnai00}%
\bibitem{Longhi04}%
  \BibitemOpen
  \bibfield{author}{%
  \bibinfo {author} {\bibfnamefont{S.}~\bibnamefont{Longhi}}, \bibinfo {author}
  {\bibfnamefont{K.}~\bibnamefont{Janner}},\ and\ \bibinfo {author}
  {\bibfnamefont{P.}~\bibnamefont{Laporta}},\ }%
  \bibfield{journal}{%
  \bibinfo {journal} {J. Opt. B}\ }%
  \textbf{\bibinfo {volume} {6}},\ \bibinfo {pages} {477} (\bibinfo {year}
  {2004})%
  \bibAnnoteFile{NoStop}{Longhi04}%
\bibitem{Novitsky06}%
  \BibitemOpen
  \bibfield{author}{%
  \bibinfo {author} {\bibfnamefont{A.~V.}\ \bibnamefont{Novitsky}}\ and\
  \bibinfo {author} {\bibfnamefont{D.~V.}\ \bibnamefont{Novitsky}},\ }%
  \bibfield{journal}{%
  \bibinfo {journal} {J. Phys. A: Math. Gen.}\ }%
  \textbf{\bibinfo {volume} {39}},\ \bibinfo {pages} {5227–} (\bibinfo {year}
  {2006})%
  \bibAnnoteFile{NoStop}{Novitsky06}%
\bibitem{Zapata08b}%
  \BibitemOpen
  \bibfield{author}{%
  \bibinfo {author} {\bibfnamefont{C.~J.}\
  \bibnamefont{Zapata-Rodr\'{\i}guez}}, \bibinfo {author}
  {\bibfnamefont{M.~A.}\ \bibnamefont{Porras}},\ and\ \bibinfo {author}
  {\bibfnamefont{J.~J.}\ \bibnamefont{Miret}},\ }%
  \bibfield{journal}{%
  \bibinfo {journal} {J. Opt. Soc. Am. A}\ }%
  \textbf{\bibinfo {volume} {25}},\ \bibinfo {pages} {2758} (\bibinfo {year}
  {2008})%
  \bibAnnoteFile{NoStop}{Zapata08b}%
\bibitem{Zapata10a}%
  \BibitemOpen
  \bibfield{author}{%
  \bibinfo {author} {\bibfnamefont{C.~J.}\
  \bibnamefont{Zapata-Rodr\'{\i}guez}}\ and\ \bibinfo {author}
  {\bibfnamefont{J.~J.}\ \bibnamefont{Miret}},\ }%
  \bibfield{journal}{%
  \bibinfo {journal} {arXiv:0910.5642v1 [physics.optics]}}%
   (\bibinfo {year} {2009})%
  \bibAnnoteFile{NoStop}{Zapata10a}%
\bibitem{Miret08}%
  \BibitemOpen
  \bibfield{author}{%
  \bibinfo {author} {\bibfnamefont{J.~J.}\ \bibnamefont{Miret}}\ and\ \bibinfo
  {author} {\bibfnamefont{C.~J.}\ \bibnamefont{Zapata-Rodr\'{\i}guez}},\ }%
  \bibfield{journal}{%
  \bibinfo {journal} {J. Opt. Soc. Am. B}\ }%
  \textbf{\bibinfo {volume} {25}},\ \bibinfo {pages} {1} (\bibinfo {year}
  {2008})%
  \bibAnnoteFile{NoStop}{Miret08}%
\bibitem{Christodoulides03}%
  \BibitemOpen
  \bibfield{author}{%
  \bibinfo {author} {\bibfnamefont{D.~N.}\ \bibnamefont{Christodoulides}},
  \bibinfo {author} {\bibfnamefont{F.}~\bibnamefont{Lederer}},\ and\ \bibinfo
  {author} {\bibfnamefont{Y.}~\bibnamefont{Silberberg}},\ }%
  \bibfield{journal}{%
  \bibinfo {journal} {Nature}\ }%
  \textbf{\bibinfo {volume} {424}},\ \bibinfo {pages} {817} (\bibinfo {year}
  {2003})%
  \bibAnnoteFile{NoStop}{Christodoulides03}%
\bibitem{Manela05}%
  \BibitemOpen
  \bibfield{author}{%
  \bibinfo {author} {\bibfnamefont{O.}~\bibnamefont{Manela}}, \bibinfo {author}
  {\bibfnamefont{M.}~\bibnamefont{Segev}},\ and\ \bibinfo {author}
  {\bibfnamefont{D.~N.}\ \bibnamefont{Christodoulides}},\ }%
  \bibfield{journal}{%
  \bibinfo {journal} {Opt. Lett.}\ }%
  \textbf{\bibinfo {volume} {30}},\ \bibinfo {pages} {2611} (\bibinfo {year}
  {2005})%
  \bibAnnoteFile{NoStop}{Manela05}%
\bibitem{Staliunas06}%
  \BibitemOpen
  \bibfield{author}{%
  \bibinfo {author} {\bibfnamefont{K.}~\bibnamefont{Staliunas}}\ and\ \bibinfo
  {author} {\bibfnamefont{R.}~\bibnamefont{Herrero}},\ }%
  \bibfield{journal}{%
  \bibinfo {journal} {Phys. Rev. E}\ }%
  \textbf{\bibinfo {volume} {73}},\ \bibinfo {pages} {016601} (\bibinfo {year}
  {2006})%
  \bibAnnoteFile{NoStop}{Staliunas06}%
\bibitem{Bouchal95b}%
  \BibitemOpen
  \bibfield{author}{%
  \bibinfo {author} {\bibfnamefont{Z.}~\bibnamefont{Bouchal}}\ and\ \bibinfo
  {author} {\bibfnamefont{M.}~\bibnamefont{Olivik}},\ }%
  \bibfield{journal}{%
  \bibinfo {journal} {J. Mod. Opt.}\ }%
  \textbf{\bibinfo {volume} {45}},\ \bibinfo {pages} {1555} (\bibinfo {year}
  {1995})%
  \bibAnnoteFile{NoStop}{Bouchal95b}%
\bibitem{Bouchal98}%
  \BibitemOpen
  \bibfield{author}{%
  \bibinfo {author} {\bibfnamefont{Z.}~\bibnamefont{Bouchal}}, \bibinfo
  {author} {\bibfnamefont{J.}~\bibnamefont{Bajer}},\ and\ \bibinfo {author}
  {\bibfnamefont{M.}~\bibnamefont{Bertolotti}},\ }%
  \bibfield{journal}{%
  \bibinfo {journal} {J. Opt. Soc. Am. A}\ }%
  \textbf{\bibinfo {volume} {15}},\ \bibinfo {pages} {2172} (\bibinfo {year}
  {1998})%
  \bibAnnoteFile{NoStop}{Bouchal98}%
\bibitem{Paakkonen02}%
  \BibitemOpen
  \bibfield{author}{%
  \bibinfo {author} {\bibfnamefont{P.}~\bibnamefont{P{\"a}{\"a}kk{\"o}nen}},
  \bibinfo {author} {\bibfnamefont{J.}~\bibnamefont{Tervo}}, \bibinfo {author}
  {\bibfnamefont{P.}~\bibnamefont{Vahimaa}}, \bibinfo {author}
  {\bibfnamefont{J.}~\bibnamefont{Turunen}},\ and\ \bibinfo {author}
  {\bibfnamefont{F.}~\bibnamefont{Gori}},\ }%
  \bibfield{journal}{%
  \bibinfo {journal} {Opt. Express}\ }%
  \textbf{\bibinfo {volume} {10}},\ \bibinfo {pages} {949} (\bibinfo {year}
  {2002})%
  \bibAnnoteFile{NoStop}{Paakkonen02}%
\bibitem{Joannopoulos08}%
  \BibitemOpen
  \bibfield{author}{%
  \bibinfo {author} {\bibfnamefont{J.~D.}\ \bibnamefont{Joannopoulos}},
  \bibinfo {author} {\bibfnamefont{S.~G.}\ \bibnamefont{Johnson}}, \bibinfo
  {author} {\bibfnamefont{J.~N.}\ \bibnamefont{Winn}},\ and\ \bibinfo {author}
  {\bibfnamefont{R.~D.}\ \bibnamefont{Meade}},\ }%
  \emph{\bibinfo {title} {Photonic crystals. Molding the flow of light}}\
  (\bibinfo {publisher} {Princeton University Press},\ \bibinfo {year} {2008})%
  \bibAnnoteFile{NoStop}{Joannopoulos08}%
\bibitem{Ramakrishna04}%
  \BibitemOpen
  \bibfield{author}{%
  \bibinfo {author} {\bibfnamefont{S.~A.}\ \bibnamefont{Ramakrishna}}\ and\
  \bibinfo {author} {\bibfnamefont{O.~J.~F.}\ \bibnamefont{Martin}},\ }%
  \bibfield{journal}{%
  \bibinfo {journal} {Opt. Lett.}\ }%
  \textbf{\bibinfo {volume} {30}},\ \bibinfo {pages} {2626} (\bibinfo {year}
  {2004})%
  \bibAnnoteFile{NoStop}{Ramakrishna04}%
\bibitem{Landau84}%
  \BibitemOpen
  \bibfield{author}{%
  \bibinfo {author} {\bibfnamefont{L.~D.}\ \bibnamefont{Landau}}, \bibinfo
  {author} {\bibfnamefont{E.~M.}\ \bibnamefont{Lifshitz}},\ and\ \bibinfo
  {author} {\bibfnamefont{L.~P.}\ \bibnamefont{Pitaevskii}},\ }%
  \emph{\bibinfo {title} {Electrodynamics of continuous media}}\ (\bibinfo
  {publisher} {Butterworth-Heinenann, Oxford, England},\ \bibinfo {year}
  {1984})%
  \bibAnnoteFile{NoStop}{Landau84}%
\bibitem{Yeh88}%
  \BibitemOpen
  \bibfield{author}{%
  \bibinfo {author} {\bibfnamefont{P.}~\bibnamefont{Yeh}},\ }%
  \emph{\bibinfo {title} {Optical Waves in Layered Media}}\ (\bibinfo
  {publisher} {Wiley},\ \bibinfo {address} {New York},\ \bibinfo {year}
  {1988})%
  \bibAnnoteFile{NoStop}{Yeh88}%
\bibitem{Kuzmiak97}%
  \BibitemOpen
  \bibfield{author}{%
  \bibinfo {author} {\bibfnamefont{V.}~\bibnamefont{Kuzmiak}}\ and\ \bibinfo
  {author} {\bibfnamefont{A.~A.}\ \bibnamefont{Maradudin}},\ }%
  \bibfield{journal}{%
  \bibinfo {journal} {Phys. Rev. B}\ }%
  \textbf{\bibinfo {volume} {55}},\ \bibinfo {pages} {7427} (\bibinfo {year}
  {1997})%
  \bibAnnoteFile{NoStop}{Kuzmiak97}%
\bibitem{Raether88}%
  \BibitemOpen
  \bibfield{author}{%
  \bibinfo {author} {\bibfnamefont{H.}~\bibnamefont{Raether}},\ }%
  \emph{\bibinfo {title} {Surface plasmons on smooth and rough surfaces and on
  gratings}}\ (\bibinfo {publisher} {Springer-Verlag},\ \bibinfo {address}
  {Berlin},\ \bibinfo {year} {1988})%
  \bibAnnoteFile{NoStop}{Raether88}%
\bibitem{Ciattoni03}%
  \BibitemOpen
  \bibfield{author}{%
  \bibinfo {author} {\bibfnamefont{A.}~\bibnamefont{Ciattoni}}\ and\ \bibinfo
  {author} {\bibfnamefont{C.}~\bibnamefont{Palma}},\ }%
  \bibfield{journal}{%
  \bibinfo {journal} {Opt. Commun.}\ }%
  \textbf{\bibinfo {volume} {224}},\ \bibinfo {pages} {175} (\bibinfo {year}
  {2003})%
  \bibAnnoteFile{NoStop}{Ciattoni03}%
\bibitem{Grunwald00}%
  \BibitemOpen
  \bibfield{author}{%
  \bibinfo {author} {\bibfnamefont{R.}~\bibnamefont{Grunwald}}, \bibinfo
  {author} {\bibfnamefont{U.}~\bibnamefont{Griebner}}, \bibinfo {author}
  {\bibfnamefont{F.}~\bibnamefont{Tschirschwitz}}, \bibinfo {author}
  {\bibfnamefont{E.~T.~J.}\ \bibnamefont{Nibbering}}, \bibinfo {author}
  {\bibfnamefont{T.}~\bibnamefont{Elsaesser}}, \bibinfo {author}
  {\bibfnamefont{V.}~\bibnamefont{Kebbel}}, \bibinfo {author}
  {\bibfnamefont{H.-J.}\ \bibnamefont{Hartmann}},\ and\ \bibinfo {author}
  {\bibfnamefont{W.}~\bibnamefont{J{\"u}ptner}},\ }%
  \bibfield{journal}{%
  \bibinfo {journal} {Opt. Lett.}\ }%
  \textbf{\bibinfo {volume} {25}},\ \bibinfo {pages} {981} (\bibinfo {year}
  {2000})%
  \bibAnnoteFile{NoStop}{Grunwald00}%
\bibitem{Zapata06e}%
  \BibitemOpen
  \bibfield{author}{%
  \bibinfo {author} {\bibfnamefont{C.~J.}\ \bibnamefont{Zapata-Rodr\'{i}guez}}\
  and\ \bibinfo {author} {\bibfnamefont{A.}~\bibnamefont{S\'anchez-Losa}},\ }%
  \bibfield{journal}{%
  \bibinfo {journal} {J. Opt. Soc. Am. A}\ }%
  \textbf{\bibinfo {volume} {23}},\ \bibinfo {pages} {3016} (\bibinfo {year}
  {2006})%
  \bibAnnoteFile{NoStop}{Zapata06e}%
\bibitem{Kuntz09}%
  \BibitemOpen
  \bibfield{author}{%
  \bibinfo {author} {\bibfnamefont{K.~B.}\ \bibnamefont{Kuntz}}, \bibinfo
  {author} {\bibfnamefont{B.}~\bibnamefont{Braverman}}, \bibinfo {author}
  {\bibfnamefont{S.~H.}\ \bibnamefont{Youn}}, \bibinfo {author}
  {\bibfnamefont{M.}~\bibnamefont{Lobino}}, \bibinfo {author}
  {\bibfnamefont{E.~M.}\ \bibnamefont{Pessina}},\ and\ \bibinfo {author}
  {\bibfnamefont{A.~I.}\ \bibnamefont{Lvovsky}},\ }%
  \bibfield{journal}{%
  \bibinfo {journal} {Phys. Rev. A}\ }%
  \textbf{\bibinfo {volume} {79}},\ \bibinfo {pages} {043802} (\bibinfo {year}
  {2009})%
  \bibAnnoteFile{NoStop}{Kuntz09}%
\bibitem{Horvath97}%
  \BibitemOpen
  \bibfield{author}{%
  \bibinfo {author} {\bibfnamefont{Z.~L.}\ \bibnamefont{Horv\'ath}}, \bibinfo
  {author} {\bibfnamefont{M.}~\bibnamefont{Erd\'elyi}}, \bibinfo {author}
  {\bibfnamefont{G.}~\bibnamefont{Szab\'o}}, \bibinfo {author}
  {\bibfnamefont{Z.}~\bibnamefont{Bor}}, \bibinfo {author}
  {\bibfnamefont{F.~K.}\ \bibnamefont{Tittel}},\ and\ \bibinfo {author}
  {\bibfnamefont{J.~R.}\ \bibnamefont{Cavallaro}},\ }%
  \bibfield{journal}{%
  \bibinfo {journal} {J. Opt. Soc. Am. A}\ }%
  \textbf{\bibinfo {volume} {14}},\ \bibinfo {pages} {3009} (\bibinfo {year}
  {1997})%
  \bibAnnoteFile{NoStop}{Horvath97}%
\bibitem{Holm98}%
  \BibitemOpen
  \bibfield{author}{%
  \bibinfo {author} {\bibfnamefont{S.}~\bibnamefont{Holm}},\ }%
  \bibfield{journal}{%
  \bibinfo {journal} {IEEE Trans. Ultrason., Ferroelect., Freq. Contr.}\ }%
  \textbf{\bibinfo {volume} {45}},\ \bibinfo {pages} {712} (\bibinfo {year}
  {1998})%
  \bibAnnoteFile{NoStop}{Holm98}%
\bibitem{Reivelt02b}%
  \BibitemOpen
  \bibfield{author}{%
  \bibinfo {author} {\bibfnamefont{K.}~\bibnamefont{Reivelt}}\ and\ \bibinfo
  {author} {\bibfnamefont{P.}~\bibnamefont{Saari}},\ }%
  \bibfield{journal}{%
  \bibinfo {journal} {Phys. Rev. E}\ }%
  \textbf{\bibinfo {volume} {65}},\ \bibinfo {pages} {046622} (\bibinfo {year}
  {2002})%
  \bibAnnoteFile{NoStop}{Reivelt02b}%
\bibitem{Reivelt02}%
  \BibitemOpen
  \bibfield{author}{%
  \bibinfo {author} {\bibfnamefont{K.}~\bibnamefont{Reivelt}}\ and\ \bibinfo
  {author} {\bibfnamefont{P.}~\bibnamefont{Saari}},\ }%
  \bibfield{journal}{%
  \bibinfo {journal} {Phys. Rev. E}\ }%
  \textbf{\bibinfo {volume} {66}},\ \bibinfo {pages} {056611} (\bibinfo {year}
  {2002})%
  \bibAnnoteFile{NoStop}{Reivelt02}%
\bibitem{Vasara89}%
  \BibitemOpen
  \bibfield{author}{%
  \bibinfo {author} {\bibfnamefont{A.}~\bibnamefont{Vasara}}, \bibinfo {author}
  {\bibfnamefont{J.}~\bibnamefont{Turunen}},\ and\ \bibinfo {author}
  {\bibfnamefont{A.~T.}\ \bibnamefont{Friberg}},\ }%
  \bibfield{journal}{%
  \bibinfo {journal} {J. Opt. Soc. Am. A}\ }%
  \textbf{\bibinfo {volume} {6}},\ \bibinfo {pages} {1748} (\bibinfo {year}
  {1989})%
  \bibAnnoteFile{NoStop}{Vasara89}%
\bibitem{Amako03}%
  \BibitemOpen
  \bibfield{author}{%
  \bibinfo {author} {\bibfnamefont{J.}~\bibnamefont{Amako}}, \bibinfo {author}
  {\bibfnamefont{D.}~\bibnamefont{Sawaki}},\ and\ \bibinfo {author}
  {\bibfnamefont{E.}~\bibnamefont{Fujii}},\ }%
  \bibfield{journal}{%
  \bibinfo {journal} {J. Opt. Soc. Am. B}\ }%
  \textbf{\bibinfo {volume} {20}},\ \bibinfo {pages} {2562} (\bibinfo {year}
  {2003})%
  \bibAnnoteFile{NoStop}{Amako03}%
\bibitem{Li09}%
  \BibitemOpen
  \bibfield{author}{%
  \bibinfo {author} {\bibfnamefont{Z.}~\bibnamefont{Li}}, \bibinfo {author}
  {\bibfnamefont{K.~B.}\ \bibnamefont{Alici}}, \bibinfo {author}
  {\bibfnamefont{H.}~\bibnamefont{Caglayan}},\ and\ \bibinfo {author}
  {\bibfnamefont{E.}~\bibnamefont{Ozbay}},\ }%
  \bibfield{journal}{%
  \bibinfo {journal} {Phys. Rev. Lett.}\ }%
  \textbf{\bibinfo {volume} {102}},\ \bibinfo {pages} {143901} (\bibinfo {year}
  {2009})%
  \bibAnnoteFile{NoStop}{Li09}%
\bibitem{Chigrin04}%
  \BibitemOpen
  \bibfield{author}{%
  \bibinfo {author} {\bibfnamefont{D.~N.}\ \bibnamefont{Chigrin}},\ }%
  \bibfield{journal}{%
  \bibinfo {journal} {Phys. Rev. E}\ }%
  \textbf{\bibinfo {volume} {70}},\ \bibinfo {pages} {056611} (\bibinfo {year}
  {2004})%
  \bibAnnoteFile{NoStop}{Chigrin04}%
\bibitem{Kosaka99}%
  \BibitemOpen
  \bibfield{author}{%
  \bibinfo {author} {\bibfnamefont{H.}~\bibnamefont{Kosaka}}, \bibinfo {author}
  {\bibfnamefont{T.}~\bibnamefont{Kawashima}}, \bibinfo {author}
  {\bibfnamefont{A.}~\bibnamefont{Tomita}}, \bibinfo {author}
  {\bibfnamefont{M.}~\bibnamefont{Notomi}}, \bibinfo {author}
  {\bibfnamefont{T.}~\bibnamefont{Tamamura}}, \bibinfo {author}
  {\bibfnamefont{T.}~\bibnamefont{Sato}},\ and\ \bibinfo {author}
  {\bibfnamefont{S.}~\bibnamefont{Kawakami}},\ }%
  \bibfield{journal}{%
  \bibinfo {journal} {Appl. Phys. Lett.}\ }%
  \textbf{\bibinfo {volume} {74}},\ \bibinfo {pages} {1212–1214} (\bibinfo
  {year} {1999})%
  \bibAnnoteFile{NoStop}{Kosaka99}%
\bibitem{Chigrin03}%
  \BibitemOpen
  \bibfield{author}{%
  \bibinfo {author} {\bibfnamefont{D.~N.}\ \bibnamefont{Chigrin}}, \bibinfo
  {author} {\bibfnamefont{S.}~\bibnamefont{Enoch}}, \bibinfo {author}
  {\bibfnamefont{C.~M.}\ \bibnamefont{Sotomayor-Torres}},\ and\ \bibinfo
  {author} {\bibfnamefont{G.}~\bibnamefont{Tayeb}},\ }%
  \bibfield{journal}{%
  \bibinfo {journal} {Opt. Express}\ }%
  \textbf{\bibinfo {volume} {11}},\ \bibinfo {pages} {1203} (\bibinfo {year}
  {2003})%
  \bibAnnoteFile{NoStop}{Chigrin03}%
\bibitem{Staliunas06b}%
  \BibitemOpen
  \bibfield{author}{%
  \bibinfo {author} {\bibfnamefont{K.}~\bibnamefont{Staliunas}}, \bibinfo
  {author} {\bibfnamefont{C.}~\bibnamefont{Serrat}}, \bibinfo {author}
  {\bibfnamefont{R.}~\bibnamefont{Herrero}}, \bibinfo {author}
  {\bibfnamefont{C.}~\bibnamefont{Cojocaru}},\ and\ \bibinfo {author}
  {\bibfnamefont{J.}~\bibnamefont{Trull}},\ }%
  \bibfield{journal}{%
  \bibinfo {journal} {Phys. Rev. E}\ }%
  \textbf{\bibinfo {volume} {74}},\ \bibinfo {pages} {016605} (\bibinfo {year}
  {2006})%
  \bibAnnoteFile{NoStop}{Staliunas06b}%
\bibitem{Martinez99}%
  \BibitemOpen
  \bibfield{author}{%
  \bibinfo {author} {\bibfnamefont{M.}~\bibnamefont{Mart\'{\i}nez-Corral}},
  \bibinfo {author} {\bibfnamefont{P.}~\bibnamefont{Andr\'es}}, \bibinfo
  {author} {\bibfnamefont{C.~J.}\ \bibnamefont{Zapata-Rodr\'{\i}guez}},\ and\
  \bibinfo {author} {\bibfnamefont{M.}~\bibnamefont{Kowalczyk}},\ }%
  \bibfield{journal}{%
  \bibinfo {journal} {Opt. Commun.}\ }%
  \textbf{\bibinfo {volume} {165}},\ \bibinfo {pages} {267} (\bibinfo {year}
  {1999})%
  \bibAnnoteFile{NoStop}{Martinez99}%
\bibitem{Kukhlevsky03}%
  \BibitemOpen
  \bibfield{author}{%
  \bibinfo {author} {\bibfnamefont{S.~V.}\ \bibnamefont{Kukhlevsky}}, \bibinfo
  {author} {\bibfnamefont{M.}~\bibnamefont{Mechler}}, \bibinfo {author}
  {\bibfnamefont{L.}~\bibnamefont{Csapo}},\ and\ \bibinfo {author}
  {\bibfnamefont{K.}~\bibnamefont{Janssens}},\ }%
  \bibfield{journal}{%
  \bibinfo {journal} {Phys. Lett. A}\ }%
  \textbf{\bibinfo {volume} {319}},\ \bibinfo {pages} {439} (\bibinfo {year}
  {2003})%
  \bibAnnoteFile{NoStop}{Kukhlevsky03}%
\bibitem{Kukhlevsky04}%
  \BibitemOpen
  \bibfield{author}{%
  \bibinfo {author} {\bibfnamefont{S.~V.}\ \bibnamefont{Kukhlevsky}}, \bibinfo
  {author} {\bibfnamefont{M.}~\bibnamefont{Mechler}}, \bibinfo {author}
  {\bibfnamefont{L.}~\bibnamefont{Csap\'o}}, \bibinfo {author}
  {\bibfnamefont{K.}~\bibnamefont{Janssens}},\ and\ \bibinfo {author}
  {\bibfnamefont{O.}~\bibnamefont{Samek}},\ }%
  \bibfield{journal}{%
  \bibinfo {journal} {Phys. Rev. B}\ }%
  \textbf{\bibinfo {volume} {70}},\ \bibinfo {pages} {195428} (\bibinfo {year}
  {2004})%
  \bibAnnoteFile{NoStop}{Kukhlevsky04}%
\bibitem{Kukhlevsky04b}%
  \BibitemOpen
  \bibfield{author}{%
  \bibinfo {author} {\bibfnamefont{S.~V.}\ \bibnamefont{Kukhlevsky}}\ and\
  \bibinfo {author} {\bibfnamefont{M.}~\bibnamefont{Mechler}},\ }%
  \bibfield{journal}{%
  \bibinfo {journal} {Opt. Commun.}\ }%
  \textbf{\bibinfo {volume} {231}},\ \bibinfo {pages} {35} (\bibinfo {year}
  {2004})%
  \bibAnnoteFile{NoStop}{Kukhlevsky04b}%
\bibitem{Nyitray08}%
  \BibitemOpen
  \bibfield{author}{%
  \bibinfo {author} {\bibfnamefont{G.}~\bibnamefont{Nyitray}}, \bibinfo
  {author} {\bibfnamefont{V.}~\bibnamefont{Mathew}},\ and\ \bibinfo {author}
  {\bibfnamefont{S.~V.}\ \bibnamefont{Kukhlevsky}},\ }%
  \bibfield{journal}{%
  \bibinfo {journal} {Opt. Commun.}\ }%
  \textbf{\bibinfo {volume} {281}},\ \bibinfo {pages} {1082–} (\bibinfo {year}
  {2008})%
  \bibAnnoteFile{NoStop}{Nyitray08}%
\end{thebibliography}
\end{document}